# Valley Phenomena in the Candidate Phase Change Material $WSe_{2(1-x)}Te_{2x}$

Sean M. Oliver[1,2], Joshua Young[3], Sergiy Krylyuk[4,5], Thomas L. Reinecke[6], Albert V. Davydov[4], Patrick M. Vora[1,2,*]

[1]*Department of Physics and Astronomy, George Mason University, Fairfax, VA, USA*

[2]*Quantum Materials Center, George Mason University, Fairfax, VA, USA*

[3]*Department of Physics, Applied Physics and Astronomy, Binghamton University, Vestal, NY, USA*

[4]*Functional Nanostructured Materials Group, National Institute of Standards and Technology, Gaithersburg, MD, USA*

[5]*Theiss Research, La Jolla, CA, USA*

[6]*Naval Research Laboratory, Washington, DC, USA*

*Author to whom correspondence should be addressed

**Email:** pvora@gmu.edu

## Abstract

Alloyed transition metal dichalcogenides provide an opportunity for coupling band engineering with valleytronic phenomena in an atomically-thin platform. However, valley properties in alloys remain largely unexplored. We investigate the valley degree of freedom in monolayer alloys of the phase change candidate material $WSe_{2(1-x)}Te_{2x}$. Low temperature Raman measurements track the alloy-induced transition from the semiconducting 1H phase of $WSe_2$ to the semimetallic $1T_d$ phase of $WTe_2$. We correlate these observations with density functional theory calculations and identify new Raman modes from W-Te vibrations in the 1H alloy phase. Photoluminescence measurements show ultra-low energy emission features that highlight alloy disorder arising from the large W-Te bond lengths. Interestingly, valley polarization and coherence in alloys survive at high Te compositions and are more robust against temperature than in $WSe_2$. These findings illustrate the persistence of valley properties in alloys with highly dissimilar parent compounds and suggest band engineering can be utilized for valleytronic devices.

## Introduction

The valley contrasting properties of monolayer semiconducting transition metal dichalcogenides (TMDs) provide the possibility of manipulating information using the valley pseudospin[1–4] in direct analogy to spintronics.[5] Devices that employ this schema for computation, commonly referred to as valleytronics, benefit from optical or electrical manipulation of the valley index, spin-valley locking, low power consumption, and the absence of Joule heating.[1,3] These advantages have stimulated vigorous investigations into the properties of two-dimensional (2D) semiconducting TMDs with an emphasis on exploring the optical interband selection rules that connect photon polarization to valley index.[1–4]

The first studies of monolayer TMDs focused solely on the photoluminescence (PL) from molybdenum disulfide ($MoS_2$),[6,7] however attention shifted to valley-dependent studies.[8,9] Soon after, monolayer tungsten diselenide ($WSe_2$) was found to be a superior TMD for valleytronics, exhibiting both exciton valley polarization and valley coherence.[10–17] Valley polarization describes the probability of an exciton created in a valley to remain there until recombination, while valley coherence quantifies the probability of an exciton to remain in a superposition state of K and K′ valleys before recombining. At low temperature, the neutral exciton ($X^0$) in bare $WSe_2$ monolayers exhibits a valley polarization ranging from 40 - 70 % depending on the laser energy.[10,12,18] The depolarization of valley excitons is believed to be governed by a combination of intervalley electron-hole exchange interactions,[9,19] phonon-assisted intervalley scattering,[8,20] and Coulomb screening of the exchange interaction.[21,22] Additionally, $WSe_2$ holds the record for the largest degree of $X^0$ valley coherence in bare monolayers (≈40 % at cryogenic temperatures).[10]

Alloying can increase the technological potential of valleytronic TMDs by combining valley contrasting properties with band engineering.[23] Furthermore, alloys of structurally distinct TMDs may also enable phase change memory technologies[24–26] by reducing the energy barrier between semiconducting and semimetallic states.[27] Few studies, however, have explored valley phenomena in alloys despite the heavy interest in this field. The first exploration of valley properties in alloys focused on $Mo_{1-x}W_xSe_2$ at 5 K and found that there was a transition from the intrinsic valley polarization of $MoSe_2$ to $WSe_2$ as the transition metal content was varied.[28] Experiments on a $WS_{0.6}Se_{1.4}$ alloy demonstrated a valley polarization of ≈31 % at 14 K, which was much lower than that of both $WS_2$ and $WSe_2$.[29] A recent study of $WS_xTe_{2-x}$ found that the room temperature valley polarization increased from 3 % in $WS_2$ to 37 % in an unspecified alloy composition.[30] The limited information regarding valley polarization in alloyed TMDs is a serious oversight as future valleytronics technologies will likely rely heavily on band engineering.

In this study, we examine the low-temperature optical properties of monolayer $WSe_{2(1-x)}Te_{2x}$, where the endpoints are the valleytronic semiconductor $1H$-$WSe_2$ and the topological semimetal $1T_d$-$WTe_2$. Prior studies of this alloy system were performed at 300 K and focused on unpolarized optical measurements[31] as well as electronic transport characterization.[32] Our study is unique in that it focuses on the impact of alloying on the valley excitons of $WSe_{2(1-x)}Te_{2x}$ monolayers encapsulated in hexagonal boron nitride (hBN). We ensure good interfacial contact in the heterostructure by cleaning with the nano-squeegee method,[33] which substantially improves the valley polarization. Raman measurements at 5 K show that increasing Te composition leads to the shifting and splitting of vibrational modes as well as the appearance of new modes unique to alloys. Polarization-resolved Raman measurements coupled with density functional theory (DFT) calculations of theoretical $1H$-$WTe_2$ phase allow us to assign these alloy-only features as resulting from W-Te vibrations in the 1H $WSe_{2(1-x)}Te_{2x}$ structure. Temperature-dependent, polarization-resolved PL measurements of the 1H-phase alloys demonstrate band gap tunability alongside the presence of new ultra-low energy emission features. We suggest these features are from deep mid-gap states that originate from the large difference in the W-Se and W-Te bond lengths. Despite the presence of structural disorder in the lattice, valley polarization is found to survive the alloying process for $x \leq 0.14$ while valley coherence is present for alloys $x \leq 0.37$, and interestingly we demonstrate that alloys have the ability to sustain these valley properties at higher temperatures than pure $WSe_2$. These findings illustrate the persistence of valley phenomena in significantly disordered alloys and point the way towards optimization of TMDs for novel phase change memories that naturally integrate with valleytronic devices.

## Results

*$WSe_{2(1-x)}Te_{2x}$ Structure and Vibrational Modes*

Interpretation of our experimental data is guided by DFT modeling of the $WSe_{2(1-x)}Te_{2x}$ phase diagram. The crystal structure of the endpoint compounds $1H$-$WSe_2$ ($x = 0$) and $1T_d$-$WTe_2$ ($x = 1$) are shown in **Figs. 1a** and **1b**. Regarding crystal structure notation, we refer to $T_d$ monolayers as being in the 1T$_d$ phase. Previously, T$_d$ and 1T' notations have been used interchangeably for T$_d$ monolayers since the only difference between these structures in the bulk was believed to be a slight shift of layers relative to one another. Since the T$_d$ structure contains three layers in its unit cell while the 1T' structure contains one layer, it was assumed that monolayers isolated from a bulk T$_d$ crystal were in the 1T' phase. We make the 1T$_d$ notation distinction in light of a recent report that has indicated, unlike the 1T' phase, inversion symmetry is broken in the monolayer T$_d$ lattice.[34] We find from our DFT calculations, in agreement with previous studies,[31,32] that the 1H phase is stable with a Te concentration of x ≤ 0.4, which we illustrate in **Fig. 1c**. For x ≥ 0.5, 1T$_d$ becomes the lowest energy phase. As a result, we expect valleytronic optical properties for the 1H alloys corresponding to x ≤ 0.4 and semimetallic behaviors for the 1T$_d$ alloys x ≥ 0.5. These predictions are consistent with our Raman and PL measurements described in detail below.

Prior to optical measurements, monolayers are encapsulated in an hBN heterostructure, as shown in **Fig. 1d**, to protect from the degradative effects due to exposure to the atmosphere and to provide a uniform dielectric environment.[35] We explore the effects of alloying on crystal structure through low-temperature Raman spectroscopy of hBN-encapsulated monolayer WSe$_{2(1-x)}$Te$_{2x}$, as shown in **Fig. 2a**. This Raman data is fit with multiple Lorentzian peaks to extract mode frequencies, which are plotted in **Fig. 2b** with alloy composition x. Examples of Lorentzian fits to Raman data for select alloys can be seen in **Supplementary Fig. S1**.

It is instructive to first examine the properties of the endpoints WSe$_2$ (x = 0) and WTe$_2$ (x = 1) that appear in blue in **Fig. 2a**. WSe$_2$, which naturally crystallizes in the 1H phase, is in the $D_{3h}$ point group and exhibits modes with $A'_1$ and $E'$ symmetries. In agreement with prior observations,[36,37] we identify eight Raman peaks in WSe$_2$ at 132 cm$^{-1}$, 223 cm$^{-1}$, 240 cm$^{-1}$, 250 cm$^{-1}$, 264 cm$^{-1}$, 351 cm$^{-1}$, 378 cm$^{-1}$, and 401 cm$^{-1}$. These modes and their symmetries are summarized in **Table 1**. The dominant feature in the WSe$_2$ Raman spectra at 250 cm$^{-1}$ (labeled peak 4 in **Figs. 2a** and **2b**) is an $A'_1$ symmetry mode where the transition metal is fixed and the chalcogens vibrate perpendicular to the basal plane.[38,39] In monolayer WSe$_2$, this mode overlaps with an $E'$ symmetry mode where the chalcogen and transition metal layers both vibrate in-plane, but out of phase.[38,39] These findings agree with prior calculations[36,38,39] as well as our own DFT modeling of the WSe$_2$ phonon band structure that predicts the 0 K frequency of the dominant $A'_1$ mode at 239 cm$^{-1}$ and the $E'$ mode at 236 cm$^{-1}$ (**Fig. 2c**). Additional monolayer WSe$_2$ modes are enhanced at 5 K, which are usually assigned as disorder-activated finite momentum phonons, combination modes, or difference modes.[36,37,40] We find that previous assignments of these features to Raman difference modes conflict with our low-temperature data. Difference modes originate from a two-phonon process where one phonon is absorbed (anti-Stokes process) and another is created (Stokes process). Since this requires the presence of a phonon before photoexcitation, its occurrence is expected to follow a Boltzmann-like temperature dependence that disappears at cryogenic temperatures.[41] The presence of these so-called difference modes in our measurements at 5 K calls this assignment into question. Thus, it is more likely that these features are combination modes for which there are many possible assignments (**Table 1**), however there is presently no explanation for the mode at 132 cm$^{-1}$.

We now discuss the Raman spectrum of WTe$_2$, which is presented in **Fig. 2a**. Our attempts to exfoliate large-area monolayers of WTe$_2$ were met with limited success, which may be due to the rapid oxidation rate of this material.[42] In contrast, we found it straightforward to achieve large-area bilayer WTe$_2$, which is known to be less susceptible to oxidation.[42] The Raman spectra of bilayer and monolayer WTe$_2$ are nearly identical,[43] which lets us safely use bilayer WTe$_2$ to discriminate between 1T$_d$- and 1H- WSe$_{2(1-x)}$Te$_{2x}$. Bilayer WTe$_2$ (T$_d$ phase) belongs to the C$_{2v}$ point group, and so only $A_1$ and $A_2$ symmetry modes can be observed.[44] Five peaks in the 70 - 425 cm$^{-1}$ range are present at 87 cm$^{-1}$, 107 cm$^{-1}$, 166 cm$^{-1}$, 218 cm$^{-1}$, and 327 cm$^{-1}$ and are labeled with letters in **Figs. 2a** and **2b**. Assignment of the mode symmetries is based on prior studies of WTe$_2$[43,45] and are included in **Table 1**. The feature at 327 cm$^{-1}$, labeled peak e, has not been previously observed, and we assign it as either a second-order overtone of the 166 cm$^{-1}$ $A_1$ mode (peak c) or a combination of the 107 cm$^{-1}$ $A_2$ and 218 cm$^{-1}$ $A_1$ modes (peaks b and d, respectively).

Low-temperature Raman measurements of the alloys (black curves of **Fig. 2a**) reveal fascinating new details regarding the vibrational modes that were not observed in Ref. 31 owing to thermal broadening at 300 K (see **Supplementary Fig. S2** for comparison of 5 K and 300 K spectra). Alloys with x ≤ 0.37 are in the 1H phase, which is further supported by the presence of PL in these samples to be discussed later, and exhibit complex mode structures in the ≈230 - 275 cm$^{-1}$ range. Polarization-resolved Raman measurements (**Supplementary Fig. S3**) reveal that the dominant WSe$_2$ $A_1'$ mode at 250 cm$^{-1}$ splits into two peaks at 244 cm$^{-1}$ and 253 cm$^{-1}$ for x = 0.04 (see red points in **Fig. 2b**). This differs from studies of WS$_{2(1-x)}$Se$_{2x}$ where this primary out-of-plane WSe$_2$ feature typically only shifts with alloying.[46,47] The splitting of out-of-plane vibrational modes with alloying, however, has been seen in $MoS_xSe_{2-x}$ monolayers[48,49] and has been carefully documented in few-layer $MoS_xSe_{2-x}$.[50] Jadczak et al. attributed the splitting of this feature in alloys to the polarization of the alloy unit cell due to the substitution of heavier chalcogens that introduce different force constants in the lattice.[50] Thus, due to the splitting of the primary $A_1'$ mode, we find that the $E'$ mode is distinguishable from the $A_1'$ mode in alloy monolayers and only slightly shifts to lower frequencies with increasing x.

Several other $WSe_2$-like vibrational modes show sensitivity to alloying for 1H-phase compositions $x \leq 0.37$. Second-order finite momentum peaks 6, 7, and 8 of **Figs. 2a** and **2b** shift to lower frequencies and broaden with increasing alloy composition. This alloy data may clarify disagreements in the literature on whether $A_1'$ or $E'$ phonons are the dominant contributor to these higher-order modes.[36,37] The band predictions of 1H-$WSe_2$ in **Fig. 2c** show both branches of the $E'$ mode that originate at 236 cm$^{-1}$ shift to lower frequencies away from the $\Gamma$ point. This behavior is opposite to that of the $A_1'$ mode, which originates at 239 cm$^{-1}$ and shifts to higher frequencies away from the $\Gamma$ point. Since peaks 6, 7, and 8 broaden asymmetrically on the lower frequency side of their centers as x is increased, this may indicate that the $E'$ mode, rather than the $A_1'$ mode, contributes to these higher-order processes. The shifting and broadening of the $WSe_2$ modes for compositions $x \leq 0.37$ indicate that alloying introduces significant disorder into the lattice but globally maintains the 1H phase.

A particularly interesting alloy-induced feature resolved in 1H samples is the Raman peak labeled $D_1$ at 191 cm$^{-1}$ in x = 0.04 (red points in **Fig. 2b**). This feature splits into the two peaks labeled $D_2$ and $D_3$ at 190 cm$^{-1}$ and 200 cm$^{-1}$, respectively, as x is increased. Polarization-resolved Raman measurements (**Supplementary Fig. S3**) indicate that these features have $A_1'$ symmetry and DFT calculations of the 1H-$WSe_2$ phonon band structure (**Fig. 2c**) show no $A_1'$ phonon modes present in this range. We therefore calculate the phonon band structure of metastable 1H-$WTe_2$ phase in **Fig. 2d**, which predicts an $A_1'$ $\Gamma$-point mode at 172 cm$^{-1}$. This mode is the closest to the observed results and so we assign the $D_1$ peak to an $A_1'$ mode arising from W-Te vibrations in the 1H $WSe_{2(1-x)}Te_{2x}$ alloys. We attribute its splitting to increasing force constant variations introduced into the lattice with noticeable Te content as discussed previously for the primary $A_1'$ mode of $WSe_2$.[50] Raman measurements reveal several other new peaks in the 1H alloys, which we label $D_4$, $D_5$, $D_6$, and D7 in **Fig. 2b**. These polarization-independent features (see **Supplementary Fig. S3**) most likely arise from either finite momentum WSe$_2$-like phonons or WSe$_2$ combination modes. We exclude difference modes based on the use of low-temperature spectroscopy as discussed previously. Possible assignments are $E'(M)$ for D6 and $E' + E'(M)$ for D7, while D4 and D5 remain unassigned but may originate from combinations of 1H-WSe$_2$ and 1H-WTe$_2$ modes. As WSe$_{2(1-x)}$Te$_{2x}$ transitions to the 1T$_d$ phase with alloying, the Raman spectra for x ≥ 0.79 show the two primary $A_1$ vibrational modes of pure WTe$_2$, labeled peaks c and d in **Figs. 2a** and **2b**. These features are shifted and broadened due to alloy disorder in agreement with other WTe$_2$ alloys.[51]

*Excitonic Properties*

In **Fig. 3**, we present temperature-dependent PL spectra for representative 1H-phase alloys. Each spectrum has been normalized by the maximum intensity and was excited with right circularly polarized light (σ+) at 1.96 eV. The collected PL is passed through a waveplate/analyzer combination to select σ+ emission. At 300 K, the PL spectra show contributions from both the neutral exciton (X$^0$) and the trion (X$^T$).[13] Increasing the Te concentration causes both features to redshift due to the lower band gap of 1H-WTe$_2$.[52] As temperature decreases from 300 K to 5 K, X$^0$ and X$^T$ sharpen, blueshift, and weaken in intensity. The latter behavior is due to the sign of the conduction band spin-orbit coupling, which makes the lowest exciton state dark (see the band schematic of WSe$_2$ in **Fig. 5a**).[53] The sign of the spin-orbit coupling is the same for all W-based TMDs[53] and so we expect no change in the optical activity of the lowest exciton state with Te substitution. The presence of broad features at lower energies compared to X$^0$ and X$^T$ originate from a combination of higher-order excitonic complexes[54] and localized exciton states from lattice defects, strain, and residual impurities introduced during fabrication.[10,13,55]

We explore band gap tunability of 1H-WSe$_{2(1-x)}$Te$_{2x}$ by extracting X$^0$ energy as a function of temperature, which is plotted in **Fig. 4a**. This data is fit with a semi-empirical formula for temperature-dependent optical band gaps given by[56]

$$E_g(T) = E_0 - S\langle\hbar\omega\rangle\left[coth\left(\frac{\langle\hbar\omega\rangle}{2kT}\right) - 1\right], \tag{1}$$

where $E_0$ is the gap at absolute zero, $S$ is a dimensionless electron-phonon coupling parameter, $\langle\hbar\omega\rangle$ is an average phonon energy, and $k$ is the Boltzmann constant. This formula models the reduction in band gap with increasing temperature due to a combination of increasing lattice constants and exciton-phonon coupling.[56] Fits of our data to equation (1) are presented as solid lines in **Fig. 4a** and the compositional dependence of $E_0$, $S$, and $\langle \hbar\omega \rangle$ from these fits are plotted in **Figs. 4b-d**. **Figure 4b** shows that $E_0$ (i.e. $X^0$ energy) varies in 1H-WSe$_{2(1-x)}$Te$_{2x}$ from ≈1.735 eV in WSe$_2$ to ≈1.519 eV in the x = 0.37 alloy. These experimental results agree extremely well with the optical band gaps of the alloys computed using DFT with the HSE06 exchange-correlation functional,[57,58] which predict a gap of 1.75 eV in 1H-phase WSe$_2$ that decreases to 1.53 eV at a Te concentration of x = 0.375 (blue triangles of **Fig. 4b**). At higher x the system transforms to the 1T$_d$ phase (**Supplementary Fig. S4**) which is a semimetal. For a comparison of DFT results for the optical band gaps of all WSe$_{2(1-x)}$Te$_{2x}$ monolayers calculated using both the HSE06 and PBE functionals and the density of states determined using the HSE06 functional, see **Supplementary Fig. S4**. Since the alloys are in the 1H phase only up to x = 0.37 and 1H-WTe$_2$ does not exist in nature, we are unable to reliably determine the bowing parameter for $E_0$.[23] Therefore, we instead fit $E_0$ from x = 0 to 1 with a linear function as shown by the red line in **Fig. 4b**. From a linear extrapolation to x = 1, we determine the 0 K optical band gap of 1H-WTe$_2$ to be 1.15 eV. Lastly, extracted values for $S$ range from 1.93 - 2.24 and for $\langle \hbar\omega \rangle$ between 4 - 16 meV (32.3 - 129 cm$^{-1}$), and are plotted versus x in **Figs. 4c** and **4d**.

Elaborating on the 5 K PL spectra of **Fig. 3**, we find that $X^T$ appears ≈30 meV below $X^0$ in our data and shifts in lockstep with the neutral exciton so that there is only weak dependence of the $X^0$-$X^T$ binding energy on x (**Supplementary Fig. S5**). Spatial PL mapping of an x = 0.33 alloy sample at 5 K indicates little variation of the intensities, energies, and linewidths of $X^0$ and $X^T$. These maps are discussed in detail in **Supplementary Note 1** and **Supplementary Fig. S6**. Additionally, we find evidence for excitonic transitions and exciton-phonon complexes at energies *above* $X^0$ known to originate from coupling between hBN and WSe$_2$ in van der Waals heterostructures.[59,60] A detailed discussion of these features is presented in **Supplementary Note 2** and **Supplementary Fig. S7**. In 5 K and 300 K reflectance measurements presented in **Supplementary Fig. S8**, both the A and B excitons are clearly visible and the valence band spin-orbit coupling is found to increase with Te incorporation.

Localized excitons are common in W-based TMDs owing to the long lifetime of the dark exciton ground states. We identify such features in all alloys and the parent compound WSe$_2$, beginning with an emission band ≈100 meV below $X^0$ hereafter referred to as L1. For all x > 0.04, L1 is accompanied by a second localized emission feature (L2) approximately ≈300 meV below $X^0$ (**Figs. 3c-e**). This feature has never been observed in TMDs, alloyed or otherwise. L1 and L2 maintain a similar energy separation with respect to $X^0$ at all nonzero x but do exhibit variations in their temperature dependent behaviors.

The new defect band L2 must originate from a disorder unique to Te-rich alloys and may be connected to the preference of $WTe_2$ to crystallize in a $1T_d$ structure. Scanning tunneling electron microscopy measurements and molecular dynamics (MD) simulations of a similar alloy system, $WS_{2-x}Te_x$,[61] indicate that Te-substitution at levels approaching 15 % can substantially modify the structure of a 1H-phase alloy. The bond lengths and lattice constants of $1H-WSe_2$ with $1H-WTe_2$ are predicted to differ by ≈7 – 8 %,[61] which also leads to a mismatch between the metal-chalcogen bond angles. The MD simulations of Ref. [61] illustrate that these internal strains can lead to the displacement of Te atoms from the expected chalcogen site for concentrations as low as 8 %. These shifts lead to the compression and stretching of neighboring hexagonal rings, and in turn result in the displacement of the native chalcogen and even the W atoms. For the $WS_2 – WTe_2$ alloys studied in Ref. 61, continuing to increase the Te doping to 15 %, 20 %, and finally 25 % increased the displacements of W, S, and Te atoms, and ultimately drove the W atoms closer together in a prelude to the 1H - $1T_d$ transition. These Te concentrations compare favorably with the values at which we observe the presence of the L2 feature (**Fig. 3**). The presence of such substantial atomic displacements and internal strains seems to be unique to TMD alloys that are mixtures between different structural phases. We therefore suggest that the internal strains driven by Te incorporation in 1H-phase alloys create a new band of defect states (i.e. the L2 feature) lower than those typically expected from chalcogen or transition metal vacancies (i.e. the L1 feature).[55]

*Valley Phenomena*

Next, we explore valley contrasting in $1H-WSe_{2(1-x)}Te_{2x}$ alloys at 5 K using 1.96 eV excitation with σ+ polarization. The schematic band structure in **Fig. 5a** illustrates the spin-valley polarized selection rules in $WSe_2$, where transitions between the valence band and the second highest conduction band are σ+ (σ-) polarized at the K (K′) point. Any emission in the opposite polarization channel is a sign of intervalley scattering. Valley polarization is determined by measuring co- and cross-polarized PL spectra in the circular basis as shown in **Fig. 5b**. The degree of valley polarization $\rho_{VP}$ for σ+ excitation is defined as $\rho_{VP} = (I_{\sigma+} - I_{\sigma-})/(I_{\sigma+} + I_{\sigma-})$, where $I_{\sigma+}$ ($I_{\sigma-}$) is the intensity of the collected light with σ+ (σ-) orientation. Valley coherence measurements are similar except measurements are carried out in a linear basis with co-polarized (∥) and cross-polarized (⊥) configurations as shown in **Fig. 5c**. The degree of valley coherence $\rho_{VC}$ is calculated similarly as $\rho_{VC} = (I_{\parallel} - I_{\perp})/(I_{\parallel} + I_{\perp})$, where $I_{\parallel}$ ($I_{\perp}$) is the intensity of the collected light that is ∥ (⊥) to the incident light.

Alloy-dependent values of $\rho_{VP}$ for $X^0$ (black squares) and $X^T$ (red circles) at 5 K are plotted in **Fig. 5d**. We find that $\rho_{VP} \approx 49\%$ for $X^0$ in WSe$_2$, which agrees with previously reported values.[10] Achieving this value required cleaning our heterostructures post-assembly using the nano-squeegeeing technique.[33] This process improved the WSe$_2$ $\rho_{VP}$ of $X^0$ by a factor of ≈1.7, thus increasing the signal from 29 % to 49 %, whereas it appeared to have very little effect on $\rho_{VC}$. This result illustrates the impact of contaminants on valley properties and a comparison of squeegeed and non-squeegeed alloys can be found in **Supplementary Fig. S9**.

As Te is substituted into the lattice, $\rho_{VP}$ first remains unchanged at x = 0.04 and then decreases to 32 % at x = 0.14 (**Fig. 5d**). For x > 0.14, no valley polarization is observed. The trion exhibits a similar trend with x; $\rho_{VP} \approx 67\%$ for $X^T$ in WSe$_2$ and is zero for x > 0.14. To the best of our knowledge, there is no systematic theory for understanding valley depolarization from alloy disorder. However, we do observe an interesting correlation between $\rho_{VP}$ and the integrated intensity ratio of $X^T / X^0$, as shown in **Supplementary Fig. S10**. The $\rho_{VP}$ $X^T / X^0$ integrated intensity ratio generally decreases with Te incorporation from ≈2.4 in WSe$_2$ to ≈0.25 in x = 0.37. While there is an initial increase in the $\rho_{VP}$ $X^T / X^0$ intensity ratio from ≈2.4 in WSe$_2$ to ≈3.8 in the x = 0.04 sample, this is most likely the result of a superior cleaned interface via the nano-squeegee method rather than an intrinsic material property. The decrease in the $\rho_{VP}$ $X^T / X^0$ intensity ratio as x is increased suggests a connection between $\rho_{VP}$ and an apparent reduction in doping with increasing Te substitution. Valley coherence is only present for neutral excitons[10] and we plot $\rho_{VC}$ versus x at 5 K in **Fig. 5e**. $\rho_{VC} \approx 20\%$ in WSe$_2$, which is slightly lower than the literature value,[10] and decreases to ≈13 % when x = 0.04 after which point it remains essentially constant until the 1H-1T$_d$ phase transition. We note that valley coherence remains finite even when valley polarization goes to zero at large x.

Temperature dependent measurements of $\rho_{VP}$ and $\rho_{VC}$ for $X^0$ in **Fig. 6** show interesting behaviors that suggest disorder can in some cases improve valley polarization and valley coherence. In all cases for the 1H-phase alloys, we find a decrease of both quantities with increasing temperature. However, alloys exhibit a different temperature dependence when compared to WSe$_2$ (x = 0). Beginning with $\rho_{VP}$ in **Fig. 6a**, we observe a sharp decrease with increasing temperature that is consistent with prior examinations of WSe$_2$ monolayers.[13,62] Surprisingly, we find that valley polarization remains large for the x = 0.04 alloy and can exceed that of WSe$_2$, reaching a maximum enhancement factor of 3.5x at 100 K. While valley polarization is overall lower for the x = 0.14 alloy, it also exceeds that of WSe$_2$ at 100 K and has the same temperature dependence as the x = 0.04 case. We note that $\rho_{VP}$ of $X^T$ also shows similar temperature and alloy dependence, which can be seen in **Supplementary Fig. S11**. This observation indicates that, counter to intuition, alloys can exhibit valley polarization that meets or exceeds that of WSe$_2$, especially at higher temperatures.

Currently, there is no systematic understanding of the impact of alloy disorder on valley polarization. The prevailing theories of valley polarization revolve around a balance between the mean exciton lifetime $\tau_x$ and the valley relaxation lifetime $\tau_v$, which are parametrized by the relationship $\rho_{VP} = \frac{\rho_0}{1+2(\gamma_v/\gamma_x)}$, where $\gamma_v = (2\tau_v)^{-1}$ is the intervalley scattering rate and $\gamma_x = (\tau_x)^{-1}$ is the exciton recombination rate.[9,21] Temperature dependence enters this equation via the scattering and recombination rates, and the valley depolarization rate for excitons is determined by a combination of electron-hole exchange interactions,[9,19] intervalley phonon scattering,[8,20] and Coulomb screening[63] of the exchange interaction. In the recent work of Ref. 21, the authors more deeply explore the temperature dependence of $\tau_v$ and find that valley depolarization at low temperatures is driven by long-range electron-hole exchange interactions, but as temperatures are increased, intervalley phonon scattering dominates. A subsequent experimental study demonstrated that the valley polarization can be enhanced over a range of temperatures by electrostatic gating, which adds additional carriers to the material that screen the electron-hole exchange interaction.[22] Therefore, we suggest that carrier doping may be a contributing factor to the enhancement of $\rho_{VP}$ in alloys at elevated temperatures, which is supported by the larger $X^T$ / $X^0$ integrated intensity ratio for x = 0.04 compared to WSe$_2$ (**Supplementary Fig. S10**). However, we cannot conclude that doping is the only factor since the $X^T$ / $X^0$ integrated intensity ratio of the

x = 0.14 alloy is less than that of WSe$_2$ but its valley polarization is larger at 100 K. Another possible explanation for the sustained valley polarization at higher temperatures may be a reduction in $\tau_x$ due to disorder.[64] We have attempted to fit our data using functional forms provided in Ref. 21, but cannot obtain unique fitting parameters since both $\tau_x$ and $\tau_v$ are complicated functions of temperature, bright-dark exciton splitting, phonon scattering rates, and exciton relaxation times.

The valley decoherence rate differs from the valley depolarization rate by its additional sensitivity to pure dephasing ($\gamma_{dep}$). This results in a different expression $\rho_{VC} = \frac{\rho_0}{1+2(\gamma_v+\gamma_{dep})/\gamma_x}$.[65,66] Unique temperature dependencies for $\rho_{VP}$ and $\rho_{VC}$ are therefore expected and observed in **Fig. 6a** and **Fig. 6b**, respectively. We again find that for alloys the valley coherence is larger than in WSe$_2$ at higher temperatures. The impact of pure dephasing on valley coherence makes it more sensitive to scattering events than valley polarization. According to Ref. 12, changes in exciton momentum due to scattering from defects yields an in-plane magnetic field that leads to depolarization and decoherence. As the frequency of impurity scattering is increased, however, the time-averaged effective magnetic field experienced by excitons due to the electron-hole exchange interaction[19] is reduced, which can act to enhance valley properties. Further studies on WSe$_{2(1-x)}$Te$_{2x}$ using temperature-dependent and time-resolved spectroscopies such as those in Refs. 12 and 21 are required to determine the respective contributions of the above decoherence and depolarization mechanisms.

**Discussion**

We have used low-temperature Raman and temperature-dependent, polarization-resolved PL spectroscopy to characterize different crystal phases spanned by monolayer WSe$_{2(1-x)}$Te$_{2x}$ alloys and explore how incorporation of Te into the WSe$_2$ lattice affects valleytronic and semiconducting properties. DFT calculations of the phonon dispersion curves for 1H-WSe$_2$ and 1T$_d$-WTe$_2$ alongside low-temperature Raman measurements allowed us to assign the vibrational modes of the WSe$_2$ and WTe$_2$ endpoint compounds. The shifting and splitting of these vibrational modes were tracked with composition x, and we found the appearance of alloy-only features resulting from W-Te vibrations in the 1H alloys that we confirm through the combination of polarization-resolved Raman measurements and DFT calculations. Temperature-dependent PL measurements were used to demonstrate band gap tunability, identify the alloy dependence of exciton and trion states, and observe a new defect-related emission feature. DFT calculations of the optical band gap in the alloys agree very well with low-temperature PL measurements when using the HSE06 functional. Polarization-resolved PL measurements show that alloys still exhibit valley polarization and coherence, and that these valley properties can be superior to those of WSe$_2$ at higher temperatures. Reflectance measurements were also used to measure the A and B excitons in select alloys, indicating that the spin-orbit splitting of the valence band can be increased with the addition of Te. This study illustrates the resilience of valley phenomena in alloys and the prospect of their application in a novel class of phase change memory technologies that also take advantage of spintronic and valleytronic information processing.

## Methods

*Crystal Growth and Structural Characterization*

WSe$_{2(1-x)}$Te$_{2x}$ alloys (x = 0…1) were grown by the chemical vapor transport (CVT) method. Appropriate amounts of W (99.9 %), Se (99.99 %), and Te (99.9 %) powders were loaded in quartz ampoules together with ≈90 mg (≈4 mg per cm$^3$ of the ampoule's volume) of TeCl$_4$ which served as a transport agent. The ampoules were then sealed under vacuum and slowly heated in a single-zone furnace until the temperature at the source and the growth zones reached 980 °C and 830 °C, respectively. After 4 days of growth, the ampoules were ice-water quenched. Crystal phases of the alloys were determined by examining powder X-ray diffraction patterns using MDI-JADE 6.5 software package.* We found that alloys with x ≤ 0.4 crystallized in the 2H phase and those with x ≥ 0.8 were in the T$_d$ phase. Results are consistent with the reports of Ref. 31. Chemical compositions were determined by the energy-dispersive X-ray spectroscopy (EDS) using a JEOL JSM-7100F field-emission scanning electron microscope (FESEM) equipped with an Oxford Instruments X-Max 80 EDS detector.

*Sample Preparation*

For optical studies, WSe$_{2(1-x)}$Te$_{2x}$ bulk crystals were mechanically exfoliated and monolayers were identified by optical contrast. Monolayers were then fully encapsulated in an hBN heterostructure (top and bottom layer) using the viscoelastic dry-stamping method on a SiO$_2$/Si substrate (90 nm oxide thickness) to protect them from the degradative effects due to exposure to the atmosphere and to provide a uniform dielectric environment.[35] We note that before encapsulation, monolayers were exposed to the atmosphere for less than one hour. To guarantee clean hBN/TMD contact, the nano-squeegee method was used with a scan line density of $\geq$ 10 nm/line and a scan speed of $\leq$ 30 µm/s as suggested by Rosenberger et al. to physically push contaminants out from in between heterostructure interfaces.[33] Results of this procedure can be seen in the atomic force microscope image of **Supplementary Fig. S6**. Here, the contaminants removed from the interfaces are gathered along both sides of the nano-squeegeed region (dark vertical lines, ≈50 nm in height above sample).

*Optical Studies*

Raman and PL measurements were performed on home-built confocal microscopes, both in backscattering geometries, that were integrated with a single close-cycle cryostat (Montana Instruments Corporation, Bozeman, MT). A 532 nm laser was used for Raman measurements since it has been shown that this excitation source can excite first and second order features,[36] whereas a 633 nm laser was used for PL measurements since it has been shown to yield a much higher degree of valley polarization than excitation with a green laser.[18] Both setups focus the excitation source through a 0.42 NA long working distance objective with 50X magnification. For Raman measurements, the laser spot was ≈2.4 µm, and the laser power density was fixed at 66 µW/µm$^2$, while for PL measurements, the laser spot was determined to be ≈2.2 µm, and the laser power density was fixed at 21 µW/µm$^2$. Collected light in both cases was directed to a 500 nm focal length spectrometer with a liquid nitrogen-cooled CCD (Princeton Instruments, Trenton, NJ). The spectrometer and camera were calibrated using a Hg-Ar atomic line source. For spectral analysis, Raman peaks were fit with Lorentzian profiles, whereas PL peaks were fit with Gaussians.

*Density Functional Theory Calculations*

Density functional theory calculations were performed using the Vienna *ab initio* Simulation Package (VASP).[67–69] Projector augmented wave (PAW) pseudopotentials[70] and the PBE exchange-correlation functional[57] were utilized. Spin-orbit coupling was included in all calculations except for the phonon band structure, which is a standard procedure.[71] It has been recently reported that there are slight differences between the 1T' and 1T$_d$ phases in monolayer

TMDs,[34] with WTe$_2$ likely forming in the 1T$_d$ phase. However, in this work we performed all calculations with the WTe$_2$ monolayer in the 1T' structure. Owing to very small differences between the 1T' phase and 1T$_d$ phase, this assumption is appropriate. Full relaxations of the lattice parameters and ionic positions were performed on monolayer WSe$_2$ and WTe$_2$ in the 1H and 1T' phases using a 32×32×1 and 32×16×1 Γ-centered k-mesh, respectively, and a 500 eV plane-wave cutoff. The phonon band structures of these compounds were computed using density functional perturbation theory (DFPT).[72] Various chalcogen-alloyed compositions of the form WSe$_{2(1-x)}$Te$_{2x}$ were created by expanding these unit cells and substituting the appropriate amount of Te with Se (or vice-versa). Full relaxations were again performed in each case, with the k-mesh scaled appropriately to the size of the unit cell. In the case when multiple substitutional anions were needed to achieve a given composition, all combinations of the position of the alloying atoms relative to each other were tested, with the lowest energy configuration considered the ground state (**Supplementary Fig. S12**). The density of states was computed using the aforementioned PBE functional, as well as the HSE06 functional,[58] with 25 % Hartree-Fock exact exchange included.

## Data Availability

The data that support the findings of this study are available from the corresponding author upon reasonable request.

## References

1. Vitale, S. A. *et al.* Valleytronics: Opportunities, Challenges, and Paths Forward. *Small* **14**, 1801483 (2018).
2. Xu, X., Yao, W., Xiao, D. & Heinz, T. F. Spin and pseudospins in layered transition metal dichalcogenides. *Nat. Phys.* **10**, 343–350 (2014).
3. Mak, K. F., Xiao, D. & Shan, J. Light–valley interactions in 2D semiconductors. *Nat. Photonics* **12**, 451–460 (2018).
4. Schaibley, J. R. *et al.* Valleytronics in 2D materials. *Nat. Rev. Mater.* **1**, (2016).
5. Žutić, I., Fabian, J. & Das Sarma, S. Spintronics: Fundamentals and applications. *Rev. Mod. Phys.* **76**, 323–410 (2004).
6. Splendiani, A. *et al.* Emerging Photoluminescence in Monolayer MoS$_2$. *Nano Lett.* **10**, 1271–1275 (2010).
7. Mak, K. F., Lee, C., Hone, J., Shan, J. & Heinz, T. F. Atomically Thin MoS$_2$: A New Direct-Gap Semiconductor. *Phys. Rev. Lett.* **105**, 136805 (2010).

8. Zeng, H., Dai, J., Yao, W., Xiao, D. & Cui, X. Valley polarization in MoS$_2$ monolayers by optical pumping. *Nat. Nanotechnol.* **7**, 490–493 (2012).

9. Mak, K. F., He, K., Shan, J. & Heinz, T. F. Control of valley polarization in monolayer MoS$_2$ by optical helicity. *Nat. Nanotechnol.* **7**, 494–498 (2012).

10. Jones, A. M. *et al.* Optical generation of excitonic valley coherence in monolayer WSe$_2$. *Nat. Nanotechnol.* **8**, 634–638 (2013).

11. Wang, G. *et al.* Valley dynamics probed through charged and neutral exciton emission in monolayer WSe$_2$. *Phys. Rev. B* **90**, 075413 (2014).

12. Hao, K. *et al.* Direct measurement of exciton valley coherence in monolayer WSe$_2$. *Nat. Phys.* **12**, 677–682 (2016).

13. Huang, J., Hoang, T. B. & Mikkelsen, M. H. Probing the origin of excitonic states in monolayer WSe$_2$. *Sci. Rep.* **6**, 22414 (2016).

14. Wang, G. *et al.* Control of Exciton Valley Coherence in Transition Metal Dichalcogenide Monolayers. *Phys. Rev. Lett.* **117**, 187401 (2016).

15. Aivazian, G. *et al.* Magnetic control of valley pseudospin in monolayer WSe$_2$. *Nat. Phys.* **11**, 148–152 (2015).

16. Chen, S.-Y. *et al.* Superior Valley Polarization and Coherence of 2s Excitons in Monolayer WSe$_2$. *Phys. Rev. Lett.* **120**, 046402 (2018).

17. Ye, Z., Sun, D. & Heinz, T. F. Optical manipulation of valley pseudospin. *Nat. Phys.* **13**, 26–29 (2016).

18. Tatsumi, Y., Ghalamkari, K. & Saito, R. Laser energy dependence of valley polarization in transition-metal dichalcogenides. *Phys. Rev. B* **94**, 235408 (2016).

19. Yu, T. & Wu, M. W. Valley depolarization due to intervalley and intravalley electron-hole exchange interactions in monolayer MoS$_2$. *Phys. Rev. B* **89**, 205303 (2014).

20. Kioseoglou, G. *et al.* Valley polarization and intervalley scattering in monolayer MoS$_2$. *Appl. Phys. Lett.* **101**, 221907 (2012).

21. Miyauchi, Y. *et al.* Evidence for line width and carrier screening effects on excitonic valley relaxation in 2D semiconductors. *Nat. Commun.* **9**, 1–10 (2018).

22. Shinokita, K. *et al.* Continuous Control and Enhancement of Excitonic Valley Polarization in Monolayer WSe$_2$ by Electrostatic Doping. *Adv. Funct. Mater.* **29**, 1900260 (2019).

23. Xie, L. M. Two-dimensional transition metal dichalcogenide alloys: Preparation, characterization and applications. *Nanoscale* **7**, 18392–18401 (2015).

24. Duerloo, K.-A. N., Li, Y. & Reed, E. J. Structural phase transitions in two-dimensional Mo- and W-dichalcogenide monolayers. *Nat. Commun.* **5**, 4214 (2014).

25. Rehn, D. A., Li, Y., Pop, E. & Reed, E. J. Theoretical potential for low energy consumption phase change memory utilizing electrostatically-induced structural phase transitions in 2D materials. *npj Comput. Mater.* **4**, 2 (2018).

26. Wang, X. *et al.* Potential 2D Materials with Phase Transitions: Structure, Synthesis, and Device Applications. *Adv. Mater.* **1804682**, 1804682 (2018).

27. Duerloo, K.-A. N. & Reed, E. J. Structural Phase Transitions by Design in Monolayer Alloys. *ACS Nano* **10**, 289–297 (2016).

28. Wang, G. *et al.* Spin-orbit engineering in transition metal dichalcogenide alloy monolayers. *Nat. Commun.* **6**, 10110 (2015).

29. Meng, Y. *et al.* Excitonic Complexes and Emerging Interlayer Electron–Phonon Coupling in BN Encapsulated Monolayer Semiconductor Alloy: $WS_{0.6}Se_{1.4}$. *Nano Lett.* **19**, 299–307 (2019).

30. Yun, S. J. *et al.* Telluriding monolayer $MoS_2$ and $WS_2$ via alkali metal scooter. *Nat. Commun.* **8**, 2163 (2017).

31. Yu, P. *et al.* Metal-Semiconductor Phase-Transition in $WSe_{2(1-x)}Te_{2x}$ Monolayer. *Adv. Mater.* **29**, 1603991 (2017).

32. Lin, J. *et al.* Anisotropic Ordering in 1T′ Molybdenum and Tungsten Ditelluride Layers Alloyed with Sulfur and Selenium. *ACS Nano* **12**, 894–901 (2018).

33. Rosenberger, M. R. *et al.* Nano-"Squeegee" for the Creation of Clean 2D Material Interfaces. *ACS Appl. Mater. Interfaces* **10**, 10379–10387 (2018).

34. Xu, S. Y. *et al.* Electrically switchable Berry curvature dipole in the monolayer topological insulator $WTe_2$. *Nat. Phys.* **14**, 900–906 (2018).

35. Castellanos-Gomez, A. *et al.* Deterministic transfer of two-dimensional materials by all-dry viscoelastic stamping. *2D Mater.* **1**, 011002 (2014).

36. del Corro, E. *et al.* Excited Excitonic States in 1L, 2L, 3L, and Bulk $WSe_2$ Observed by Resonant Raman Spectroscopy. *ACS Nano* **8**, 9629–9635 (2014).

37. Zhao, W. *et al.* Lattice dynamics in mono- and few-layer sheets of $WS_2$ and $WSe_2$. *Nanoscale* **5**, 9677 (2013).

38. Terrones, H. *et al.* New First Order Raman-active Modes in Few Layered Transition Metal Dichalcogenides. *Sci. Rep.* **4**, 4215 (2015).

39. Luo, X. *et al.* Effects of lower symmetry and dimensionality on Raman spectra in two-dimensional $WSe_2$. *Phys. Rev. B - Condens. Matter Mater. Phys.* **88**, 1–7 (2013).

40. Sun, H. *et al.* Enhanced exciton emission behavior and tunable band gap of ternary $W(S_xSe_{1-x})_2$ monolayer: temperature dependent optical evidence and first-principles calculations. *Nanoscale* **10**, 11553–11563 (2018).

41. Cuscó, R. *et al.* Temperature dependence of Raman scattering in ZnO. *Phys. Rev. B* **75**, 165202 (2007).

42. Ye, F. *et al.* Environmental Instability and Degradation of Single- and Few-Layer $WTe_2$ Nanosheets in Ambient Conditions. *Small* **12**, 5802–5808 (2016).

43. Jiang, Y. C., Gao, J. & Wang, L. Raman fingerprint for semi-metal $WTe_2$ evolving from bulk to monolayer. *Sci. Rep.* **6**, 19624 (2016).

44. Cao, Y. *et al.* Anomalous vibrational modes in few layer $WTe_2$ revealed by polarized Raman scattering and first-principles calculations. *2D Mater.* **4**, 035024 (2017).

45. Kim, Y. *et al.* Anomalous Raman scattering and lattice dynamics in mono- and few-layer $WTe_2$. *Nanoscale* **8**, 2309–2316 (2016).

46. Duan, X. *et al.* Synthesis of $WS_{2x}Se_{2-2x}$ Alloy Nanosheets with Composition-Tunable Electronic Properties. *Nano Lett.* **16**, 264–269 (2016).

47. Fu, Q. *et al.* Synthesis and Enhanced Electrochemical Catalytic Performance of Monolayer $WS_{2(1-x)}Se_{2x}$ with a Tunable Band Gap. *Adv. Mater.* **27**, 4732–4738 (2015).

48. Mann, J. *et al.* 2-Dimensional Transition Metal Dichalcogenides with Tunable Direct Band Gaps: $MoS_{2(1-x)}Se_{2x}$ Monolayers. *Adv. Mater.* **26**, 1399–1404 (2014).

49. Feng, Q. *et al.* Growth of $MoS_{2(1-x)}Se_{2x}$ (x = 0.41 – 1.00) Monolayer Alloys with Controlled Morphology by Physical Vapor Deposition. *ACS Nano* **9**, 7450–7455 (2015).

50. Jadczak, J. *et al.* Composition dependent lattice dynamics in $MoS_xSe_{(2-x)}$ alloys. *J. Appl. Phys.* **116**, (2014).

51. Oliver, S. M. *et al.* The structural phases and vibrational properties of $Mo_{1-x}W_xTe_2$ alloys. *2D Mater.* **4**, 045008 (2017).

52. Kang, J., Tongay, S., Zhou, J., Li, J. & Wu, J. Band offsets and heterostructures of two-dimensional semiconductors. *Appl. Phys. Lett.* **102**, 012111 (2013).

53. Echeverry, J. P., Urbaszek, B., Amand, T., Marie, X. & Gerber, I. C. Splitting between bright and dark excitons in transition metal dichalcogenide monolayers. *Phys. Rev. B* **93**, 1–5 (2016).

54. Paur, M. *et al.* Electroluminescence from multi-particle exciton complexes in transition metal dichalcogenide semiconductors. *Nat. Commun.* **10**, 1709 (2019).

55. Tongay, S. *et al.* Defects activated photoluminescence in two-dimensional semiconductors: interplay between bound, charged and free excitons. *Sci. Rep.* **3**, 2657 (2013).

56. O'Donnell, K. P. & Chen, X. Temperature dependence of semiconductor band gaps. *Appl. Phys. Lett.* **58**, 2924–2926 (1991).

57. Perdew, J. P., Burke, K. & Ernzerhof, M. Generalized Gradient Approximation Made Simple. *Phys. Rev. Lett.* **77**, 3865–3868 (1996).

58. Heyd, J., Scuseria, G. E. & Ernzerhof, M. Hybrid functionals based on a screened Coulomb potential. *J. Chem. Phys.* **118**, 8207–8215 (2003).

59. Jin, C. *et al.* Interlayer electron–phonon coupling in $WSe_2$/hBN heterostructures. *Nat. Phys.* **13**, 127–131 (2017).

60. Chow, C. M. *et al.* Unusual Exciton–Phonon Interactions at van der Waals Engineered Interfaces. *Nano Lett.* **17**, 1194–1199 (2017).

61. Tang, B. *et al.* Phase-Controlled Synthesis of Monolayer Ternary Telluride with a Random Local Displacement of Tellurium Atoms. *Adv. Mater.* **1900862**, 1900862 (2019).

62. Hanbicki, A. T. *et al.* Optical polarization of excitons and trions under continuous and pulsed excitation in single layers of $WSe_2$. *Nanoscale* **9**, 17422–17428 (2017).

63. Konabe, S. Screening effects due to carrier doping on valley relaxation in transition metal dichalcogenide monolayers. *Appl. Phys. Lett.* **109**, 073104 (2016).

64. McCreary, K. M., Currie, M., Hanbicki, A. T., Chuang, H. J. & Jonker, B. T. Understanding Variations in Circularly Polarized Photoluminescence in Monolayer Transition Metal Dichalcogenides. *ACS Nano* **11**, 7988–7994 (2017).

65. Chakraborty, C., Mukherjee, A., Qiu, L. & Vamivakas, A. N. Electrically tunable valley polarization and valley coherence in monolayer $WSe_2$ embedded in a van der Waals heterostructure. *Opt. Mater. Express* **9**, 1479 (2019).

66. Qiu, L., Chakraborty, C., Dhara, S. & Vamivakas, A. N. Room-temperature valley coherence in a polaritonic system. *Nat. Commun.* **10**, 1–5 (2019).

67. Kresse, G. & Furthmüller, J. Efficiency of ab-initio total energy calculations for metals and semiconductors using a plane-wave basis set. *Comput. Mater. Sci.* **6**, 15–50 (1996).

68. Kresse, G. & Furthmüller, J. Efficient iterative schemes for ab initio total-energy calculations using a plane-wave basis set. *Phys. Rev. B - Condens. Matter Mater. Phys.* **54**, 11169–11186 (1996).

69. Kohn, W. & Sham, L. J. Self-Consistent Equations Including Exchange and Correlation Effects. *Phys. Rev.* **140**, A1133–A1138 (1965).

70. P.E., B. Projector augmented-wave method. *Phys. Rev. B* **50**, 17953–17979 (1994).

71. Xiang, H. *et al.* Quantum spin Hall insulator phase in monolayer $WTe_2$ by uniaxial strain. *AIP Adv.* **6**, 095005 (2016).

72. Baroni, S., Giannozzi, P. & Testa, A. Green's-function approach to linear response in solids. *Phys. Rev. Lett.* **58**, 1861–1864 (1987).

## Acknowledgements

P.M.V. and S.M.O. acknowledge support from the National Science Foundation (NSF) under Grant No. DMR-1748650, the George Mason University Quantum Materials Center, and the George Mason University Presidential Scholars Program. J.Y. was supported by funds from Binghamton University. Computational resources were provided by the Department of Defense High Performance Computing Modernization Program. S.K. acknowledges support from the US Department of Commerce, NIST under financial assistance award 70NANB18H155. T.L.R. was supported by the Office of Naval Research.

***Disclaimer:** Certain commercial equipment, instruments, or materials are identified in this paper in order to specify the experimental procedure adequately. Such identification is not intended to imply recommendation or endorsement by the National Institute of Standards and Technology, nor is it intended to imply that the materials or equipment identified are necessarily the best available for the purpose.

## Author Contributions

P.M.V. and S.M.O. conceived and designed the experiments and wrote the manuscript with input from all coauthors. S.M.O. prepared van der Waals heterostructures and carried out all experiments. A.D.V. and S.K. grew the bulk crystals used in this work. J.Y. and T.L.R. performed the density functional theory calculations.

## Competing Interests

The authors declare no competing financial interests.

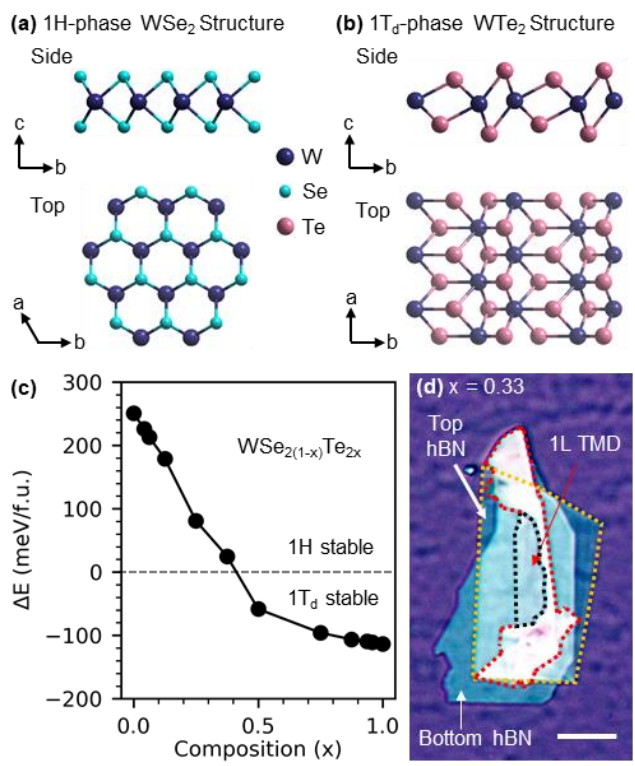

**Fig. 1 | Structural phases and van der Waals heterostructures of WSe$_{2(1-x)}$Te$_{2x}$.** Side and top view of (**a**) monolayer 1H-WSe$_2$ and (**b**) monolayer 1T$_d$-WTe$_2$. (**c**) Composition-dependent phase diagram determined from DFT calculations indicating a phase boundary at x = 0.4. (**d**) Optical image of an hBN-encapsulated x = 0.33 monolayer deposited onto a SiO$_2$/Si substrate. The TMD alloy is outlined in red and the monolayer (1L) portion of that flake is outlined in black. The top layer of hBN is outlined in yellow. The white scale bar is 20 µm.

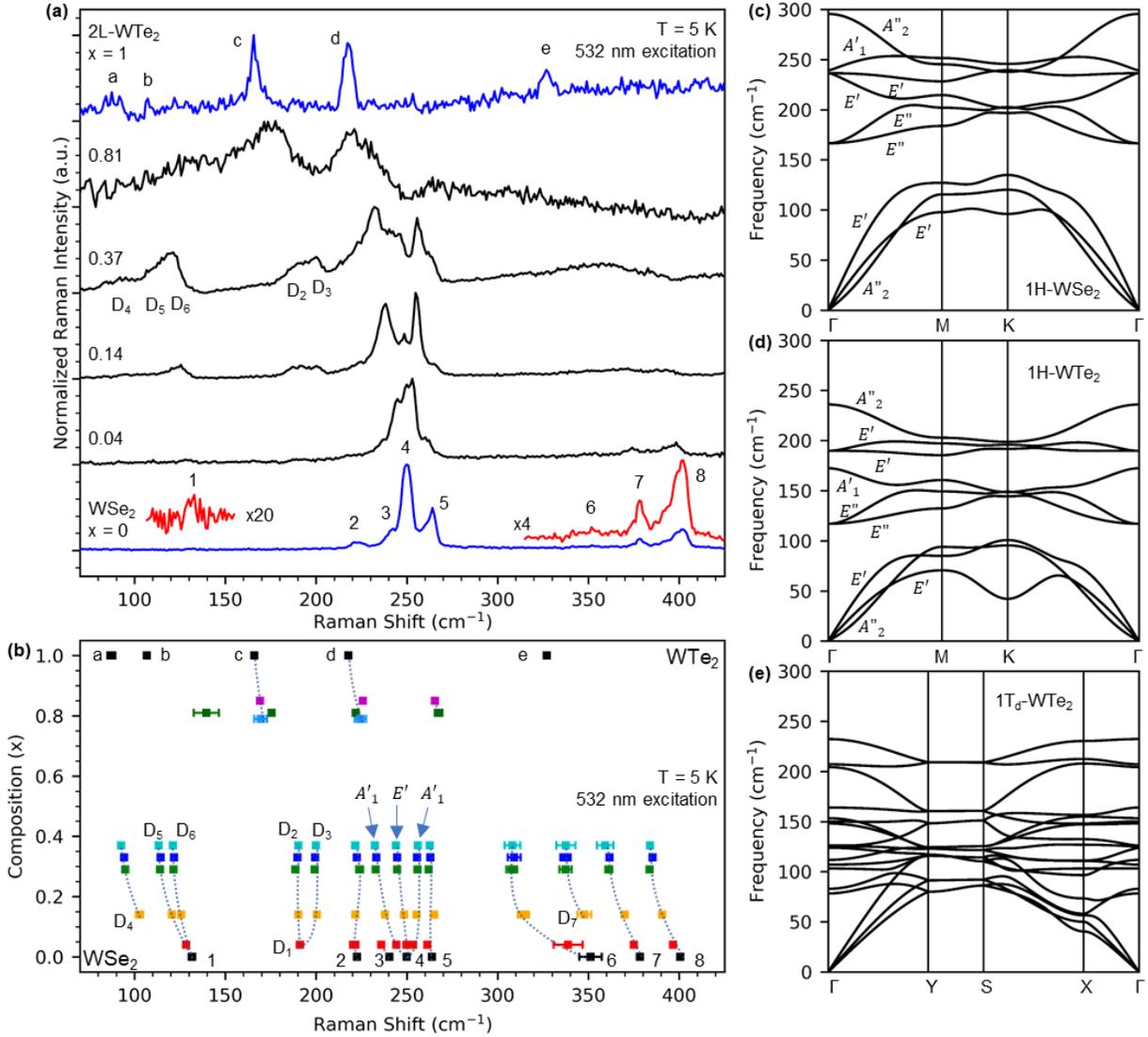

**Fig. 2 | Composition-dependent Raman spectra of WSe$_{2(1-x)}$Te$_{2x}$ and phonon band structures of WSe$_2$ and WTe$_2$. (a)** Raman measurements of monolayer WSe$_2$, bilayer (2L) WTe$_2$, and select monolayer WSe$_{2(1-x)}$Te$_{2x}$ alloys taken at 5 K with 532 nm excitation. Vibrational modes in WSe$_2$ and WTe$_2$ are identified with numbers and letters, respectively, and their assignments can be found in **Table 1**. Parts of the WSe$_2$ spectra are scaled for clarity. **(b)** Peak positions extracted from Raman measurements at 5 K for different alloy compositions x. Peaks identified in panel **(a)** for WSe$_2$ and WTe$_2$ are labeled with their respective numbers and letters. New alloy-induced vibrational modes are labeled D$_i$ (i = 1, 2, 3, …). The composition-dependent shifting and the splitting of peaks are tracked with dotted lines. The error bars in panel **(b)** are equal to one standard deviation. **(c-e)** Phonon band structures calculated for 1H-WSe$_2$, 1H-WTe$_2$, and 1T$_d$-WTe$_2$.

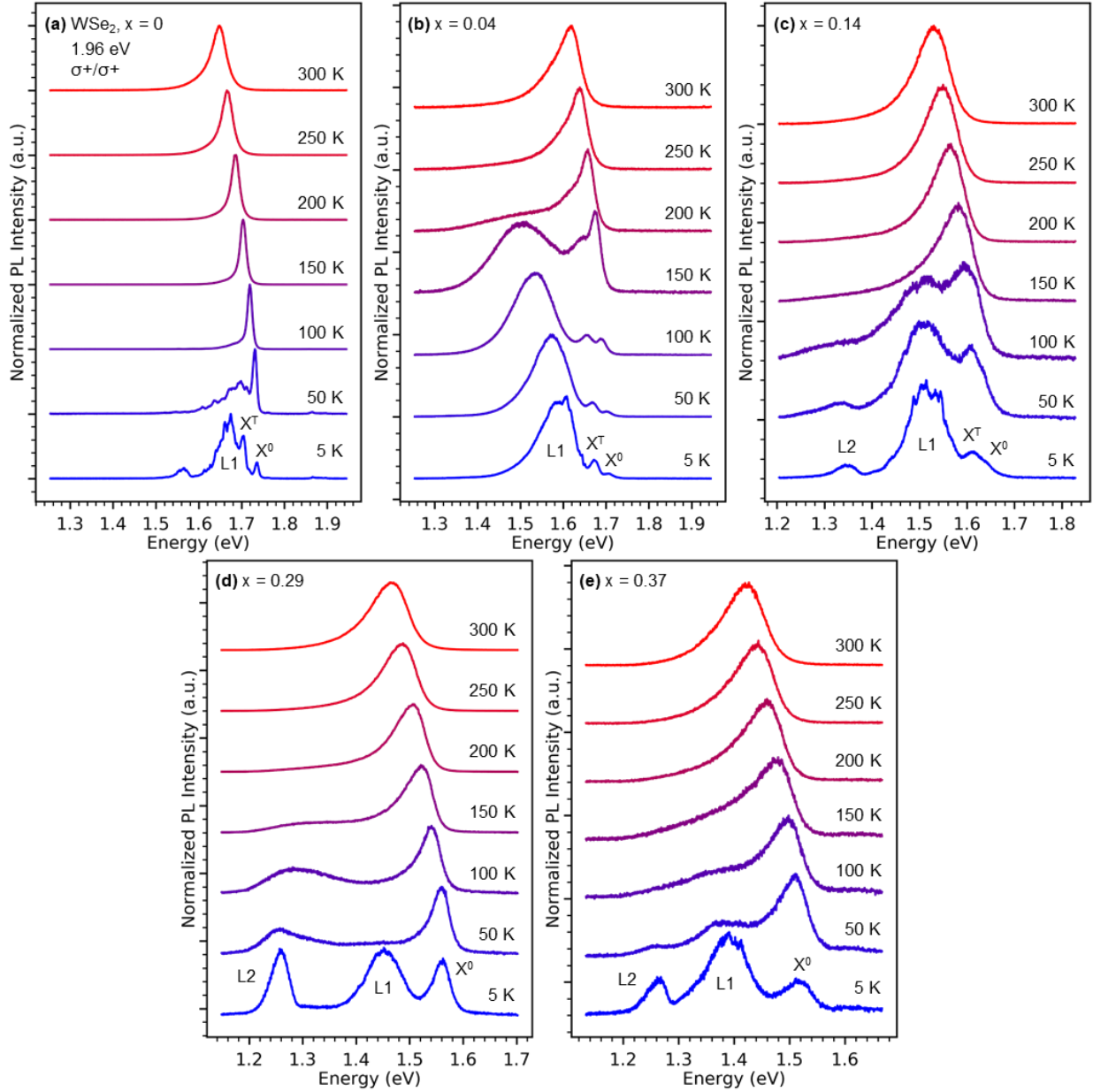

**Fig. 3 | Temperature-dependent PL of 1H-WSe$_{2(1-x)}$Te$_{2x}$.** PL measurements (1.96 eV excitation) of **(a)** WSe$_2$, as well as 1H-phase WSe$_{2(1-x)}$Te$_{2x}$ alloys corresponding to **(b)** x = 0.04, **(c)** x = 0.14, **(d)** x = 0.29, and **(e)** x = 0.37. Excitation and collection are done with right circularly polarized light (σ+). The neutral exciton (X$^0$) and trion (X$^T$) are labeled where appropriate. Emission at 300 K is dominated by X$^0$, which has a low-energy tail resulting from the presence of X$^T$. As the temperature is decreased, X$^0$ and X$^T$ sharpen and blueshift while the localized exciton features L1 and L2 begin to dominate the spectra.

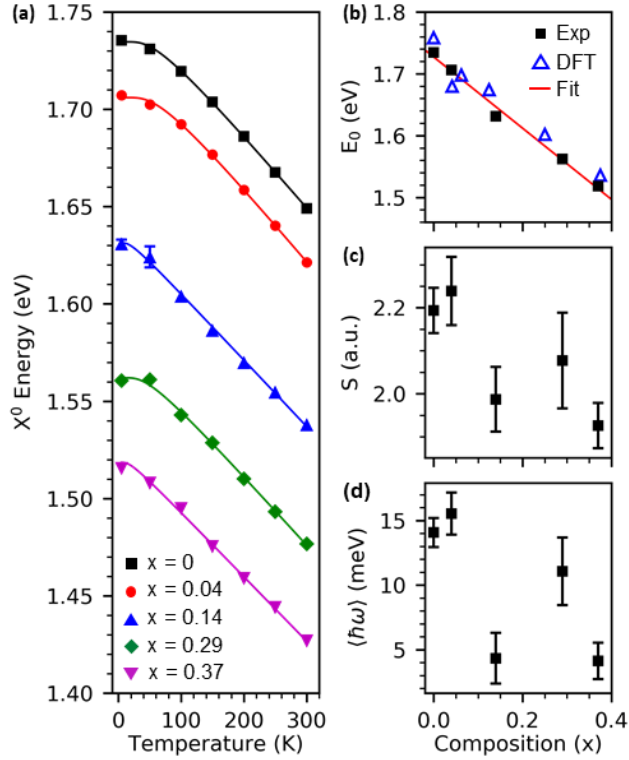

**Fig. 4 | Composition- and temperature-dependent exciton energies. (a)** $X^0$ energies extracted from temperature-dependent PL measurements (1.96 eV excitation). Excitation and collection are done with right circularly polarized light. The solid lines are fits to equation (1). The compositional dependence of the extracted parameters $E_0$, $S$, and $\langle \hbar\omega \rangle$ are plotted in panels **(b), (c),** and **(d)**, respectively. $E_0$ is found to be tunable with alloying, while $S$ and $\langle \hbar\omega \rangle$ are found to decrease with increasing alloy composition x. In panel **(b)**, we plot DFT predicted optical band gaps as blue triangles, while the red curve is a fit to a line used to extract a 0 K band gap for 1H-WTe$_2$ of 1.15 eV. Error bars shown in all panels are equal to one standard deviation.

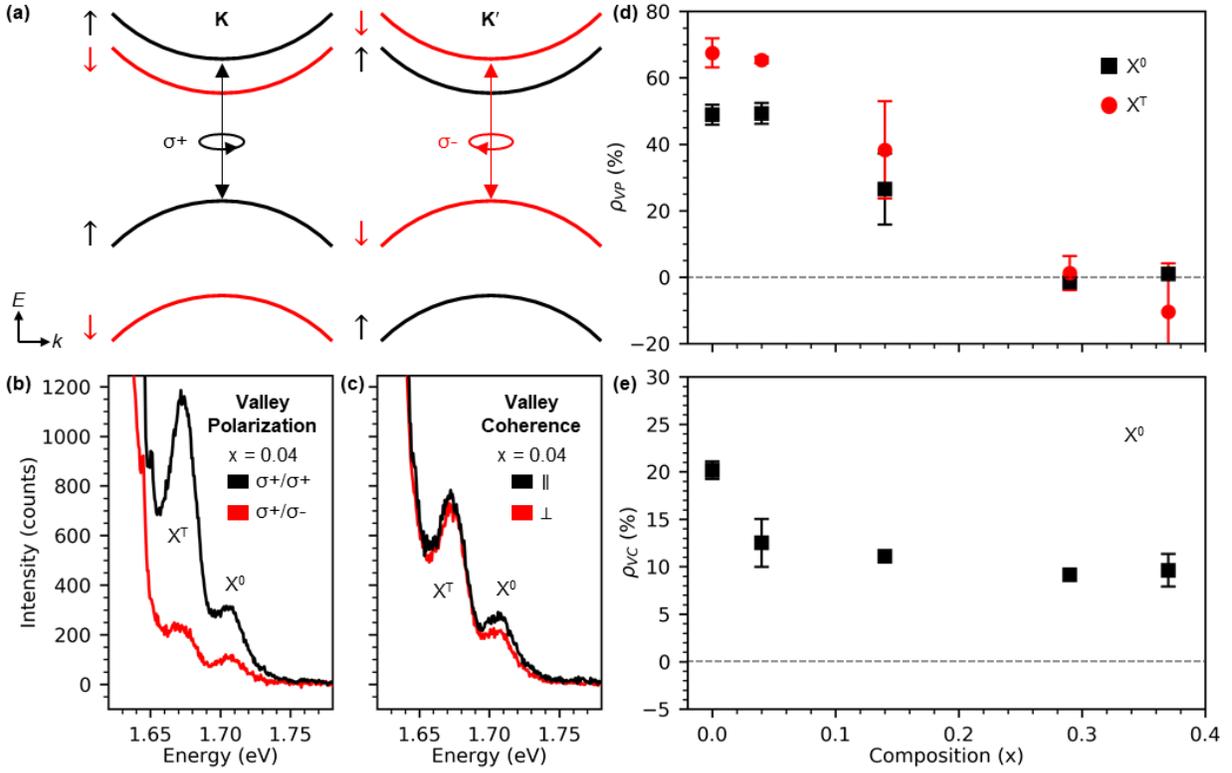

**Fig. 5 | Valley properties of WSe$_{2(1-x)}$Te$_{2x}$ at 5 K.** (**a**) Simplified image of the electron bands near the K and K′ points of the hexagonal Brillouin zone in monolayer WSe$_2$. Valley-dependent optical selection rules couple transitions at the K (K′) valleys with σ+ (σ-) circularly polarized light. (**b**) and (**c**) show example spectra of valley polarization and valley coherence measurements for x = 0.04, respectively. (**d**) The degree of valley polarization ($\rho_{VP}$) of X$^0$ (black squares) and X$^T$ (red circles) and (**e**) the degree of valley coherence ($\rho_{VC}$) of X$^0$ plotted against alloy composition x. The error bars shown in panels (**d**) and (**e**) are equal to one standard deviation.

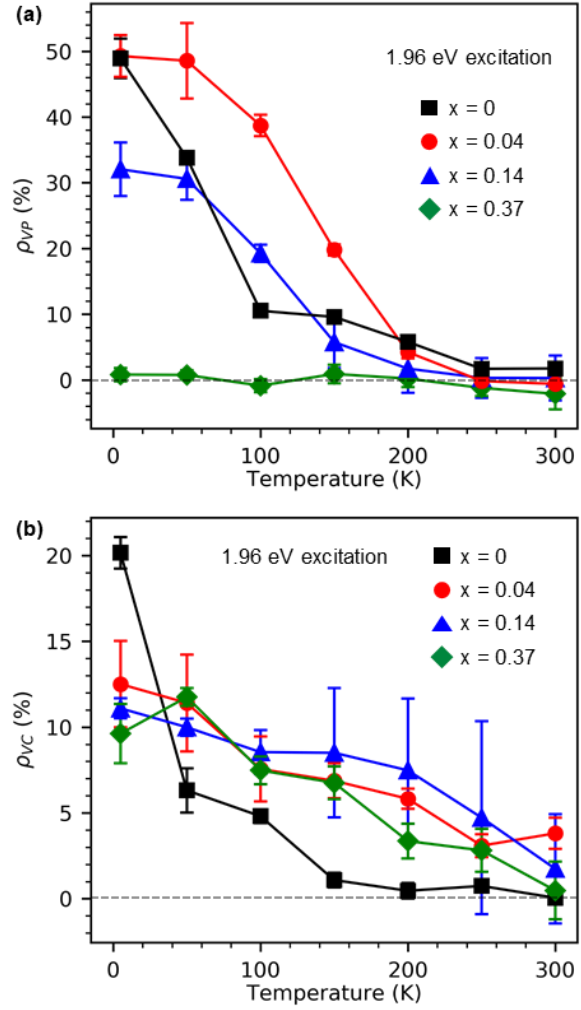

**Fig. 6 | Temperature dependence of $\rho_{VP}$ and $\rho_{VC}$ in 1H-WSe$_{2(1-x)}$Te$_{2x}$.** (a) $\rho_{VP}$ and (b) $\rho_{VC}$ of 1H-WSe$_{2(1-x)}$Te$_{2x}$ as a function of temperature. In both cases, the alloys are found to sustain valley properties at elevated temperatures when compared to pure WSe$_2$ (x = 0). Measurements in both panels are done with 1.96 eV excitation. The solid lines in both panels are guides to the eye and the error bars are equal to one standard deviation.

| 1H-WSe$_2$ | | |
|---|---|---|
| Label | Position (cm$^{-1}$) | Assignment |
| 1 | 132 | unknown |
| 2 | 223 | $E'(K)^a$ |
| 3 | 240 | $E'(M)^a$ |
| 4 | 250 | $A'_1 + E'$ |
| 5 | 264 | $2LA(M)^{a,b}$ |
| 6 | 351 | $2E'(\Gamma)^b$ or $A'_1(M) + TA(M)^b$ |
| 7 | 378 | $[E'(\Gamma)$ or $A'_1(\Gamma)] + LA(M)^{a,b}$ |
| 8 | 401 | $[E'(\Gamma)$ or $A'_1(\Gamma)] + LA(K)^a$ or $3LA(M)^{a,b}$ |

| T$_d$-WTe$_2$ (bilayer) | | |
|---|---|---|
| Label | Position (cm$^{-1}$) | Assignment |
| a | 87 | $A_2$ |
| b | 107 | $A_2$ |
| c | 166 | $A_1$ |
| d | 218 | $A_1$ |
| e | 327 | $2A_1$ (2x peak c) or $A_1 + A_2$ (peak d + peak b) |

**Table 1. | Raman mode assignments.** 1H-WSe$_2$ and T$_d$-WTe$_2$ (bilayer) vibrational mode symmetry assignments for the peaks identified in **Figs. 2a** and **2b**. WSe$_2$ peaks are labeled with numbers and WTe$_2$ peaks are labeled with letters. The superscripts a and b refer to assignments made in Refs. 36 and 37, respectively.

# Supplementary Information

## Valley Phenomena in the Candidate Phase Change Material WSe$_{2(1-x)}$Te$_{2x}$

Sean M. Oliver[1,2], Joshua Young[3], Sergiy Krylyuk[4,5], Thomas L. Reinecke[6], Albert V. Davydov[4], Patrick M. Vora[1,2,*]

[1]*Department of Physics and Astronomy, George Mason University, Fairfax, VA, United States of America*

[2]*Quantum Materials Center, George Mason University, Fairfax, VA, United States of America*

[3]*Department of Physics, Applied Physics and Astronomy, Binghamton University, Vestal, NY, United States of America*

[4]*Functional Nanostructured Materials Group, National Institute of Standards and Technology, Gaithersburg, MD, United States of America*

[5]*Theiss Research, La Jolla, CA, USA*

[6]*Naval Research Laboratory, Washington, DC, United States of America*

*Author to whom correspondence should be addressed

**Email:** pvora@gmu.edu

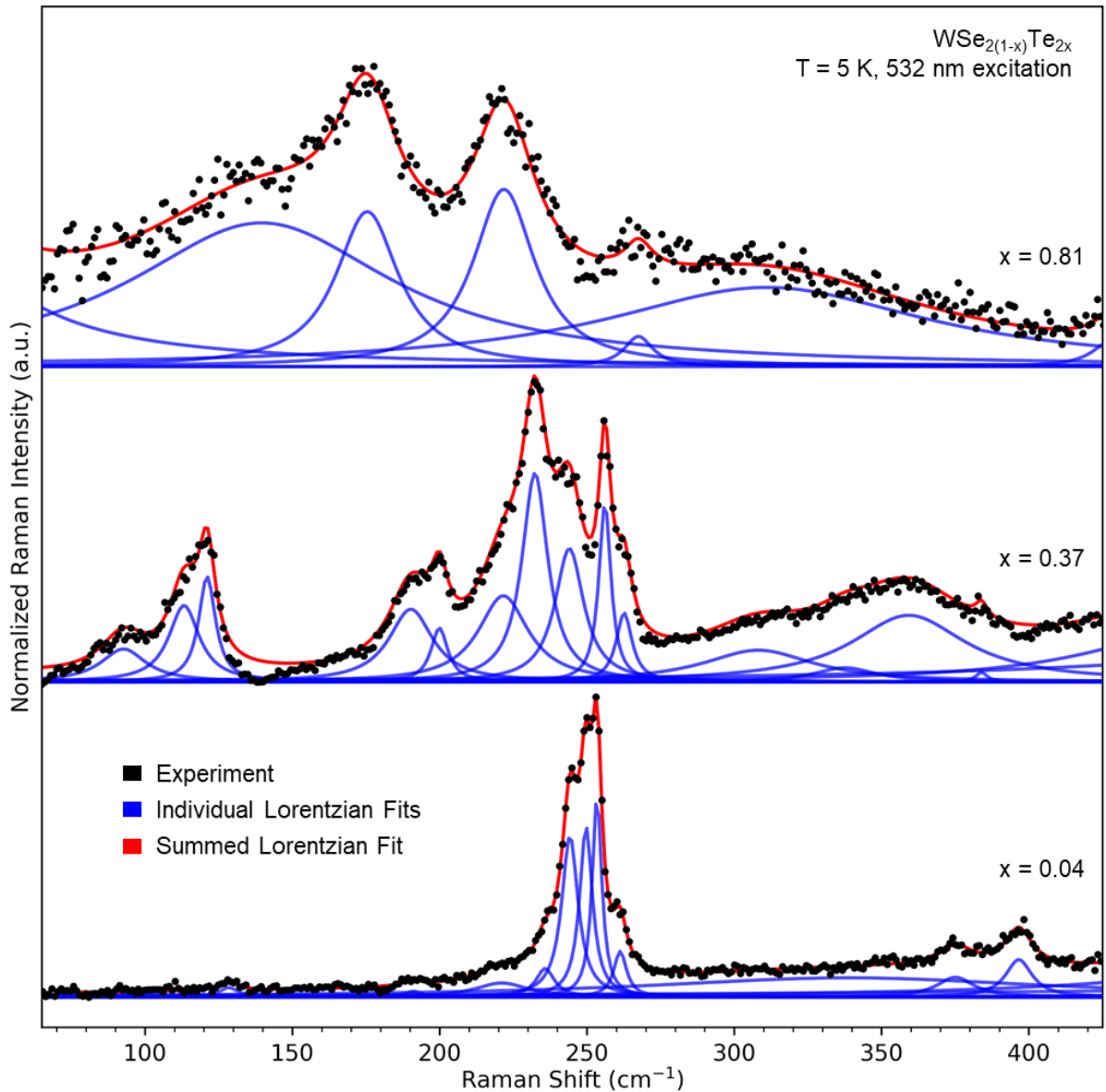

**Supplementary Figure 1 | Lorentzian fits to Raman data of select $WSe_{2(1-x)}Te_{2x}$ alloys.** Raman measurements of x = 0.04, 0.37, and 0.81 alloys taken at 5 K with 532 nm excitation. The black points correspond to the experimental data and the blue peaks are individual Lorentzian fits. The red curves are obtained by summing the individual Lorentzian fits.

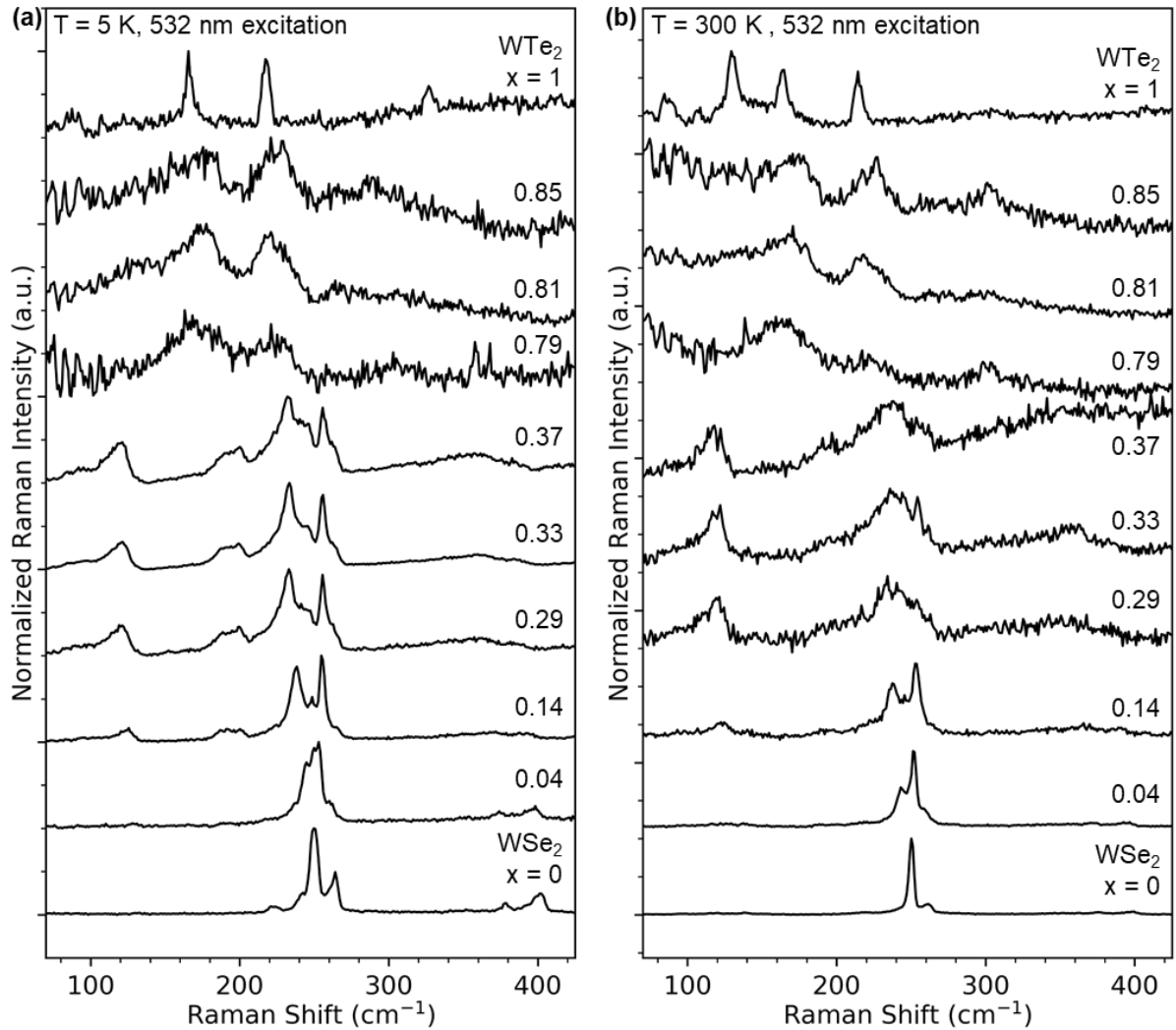

**Supplementary Figure 2 | Composition-dependent Raman measurements of $WSe_{2(1-x)}Te_{2x}$.** Raman measurements taken with 532 nm excitation at **(a)** 5 K and **(b)** 300 K for all studied compositions. We note that even though the $WTe_2$ spectra of panels **(a)** and **(b)** were taken on the same sample, the spectrum of panel **(a)** was taken on a bilayer region (absence of peak at ~130 cm$^{-1}$) while the spectrum of panel **(b)** captured a nearby portion of bulk $WTe_2$ (presence of peak at ~130 cm$^{-1}$).

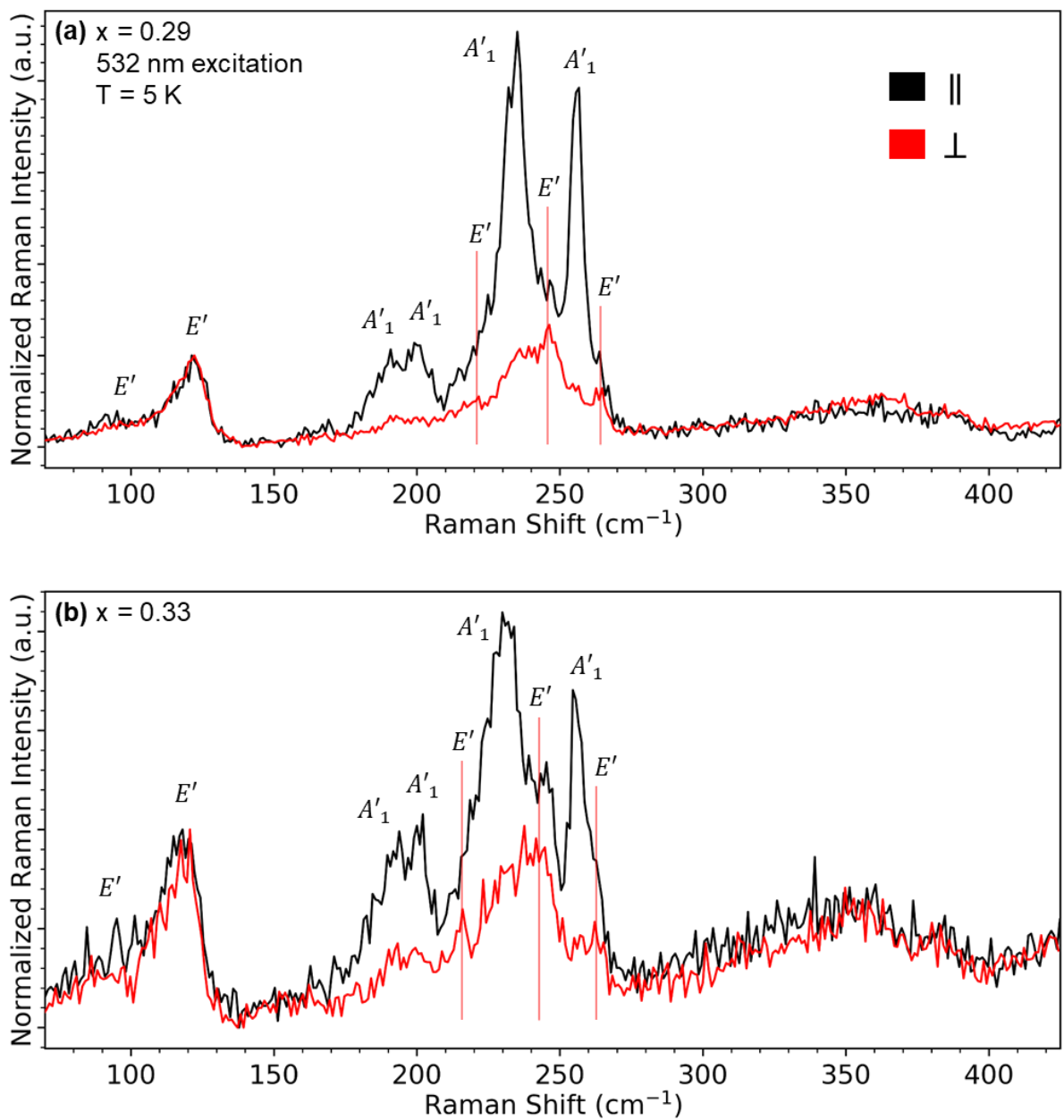

**Supplementary Figure 3 | Polarization-resolved Raman measurements of select 1H-$WSe_{2(1-x)}Te_{2x}$ alloys taken at 5 K.** Raman measurements taken with 532 nm excitation of **(a)** x = 0.29 and **(b)** x = 0.33 alloys, which were chosen for the clear splitting of the primary $A'_1$ mode in $WSe_2$. Due to the differing symmetries of the $A'_1$ and $E'$ modes, polarization selection rules in a backscattering experimental geometry state that $A'_1$ modes should only be visible when the analyzer is co-polarized with the excitation, while $E'$ modes are visible when the analyzer is arranged either co- or cross-polarized with the excitation.

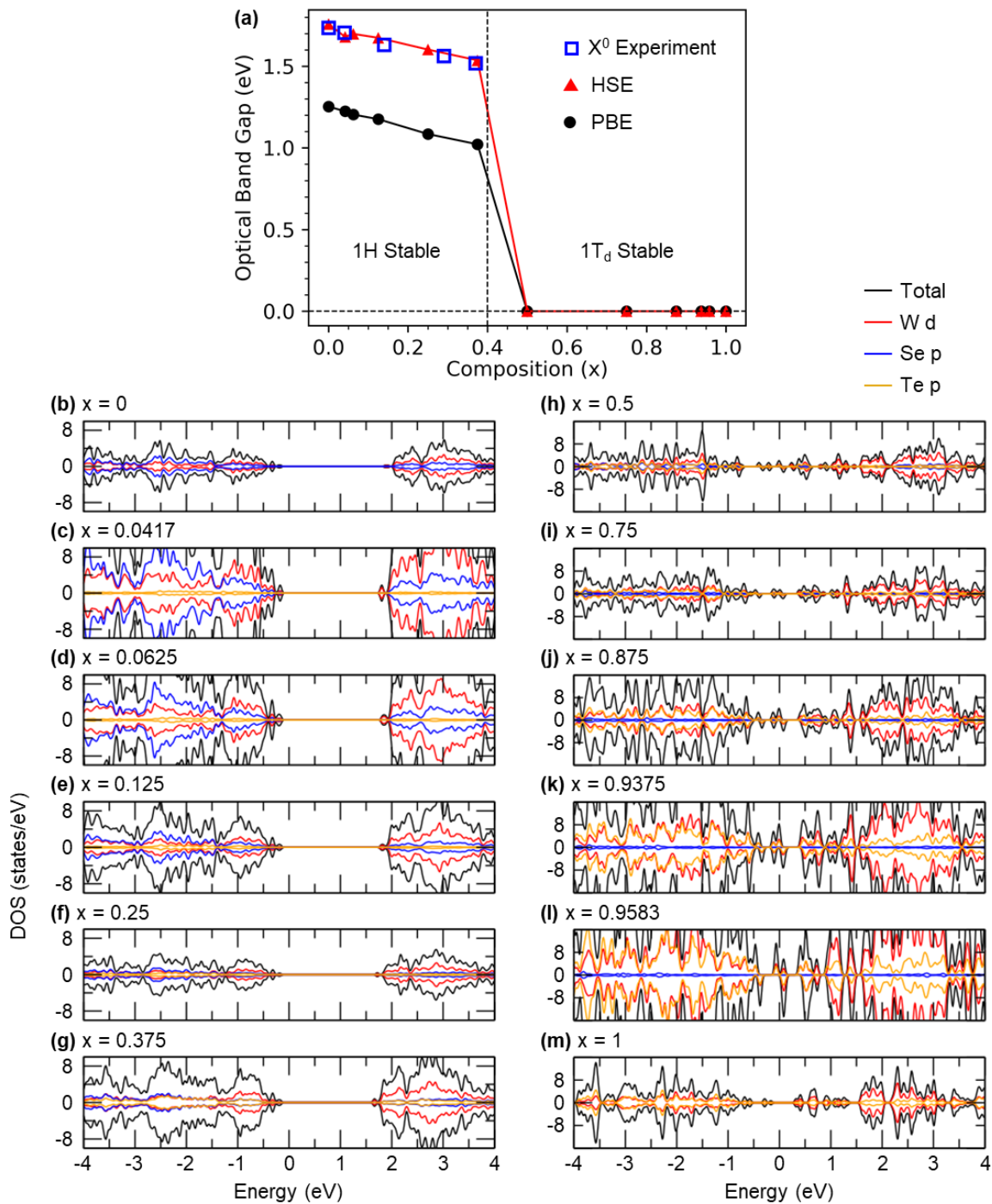

**Supplementary Figure 4 | DFT-calculated optical band gaps and density of states for WSe$_{2(1-x)}$Te$_{2x}$.** **(a)** Optical band gaps calculated using DFT. We find that the HSE06 functional predicts a band gap that agrees better with experiment than the PBE functional as demonstrated in several other material families.[1,2] **(b-m)** Density of states calculated using the HSE06 functional for monolayer WSe$_{2(1-x)}$Te$_{2x}$ alloys.

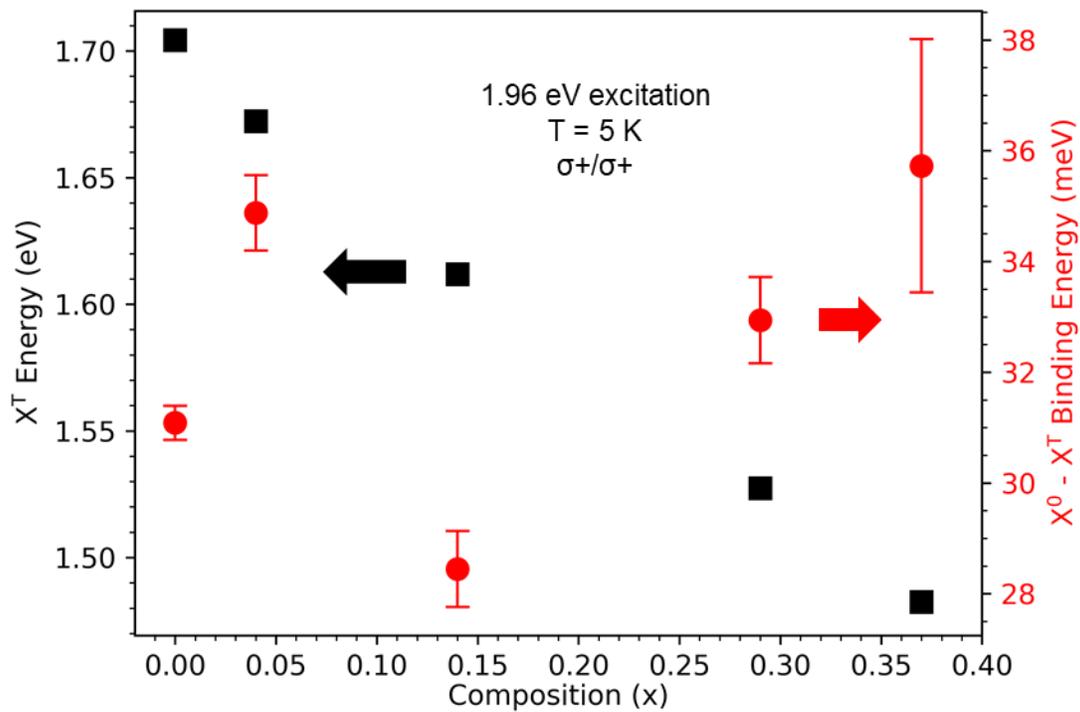

**Supplementary Figure 5 | Composition-dependent $X^T$ energy and $X^0$-$X^T$ binding energy.** The alloy dependence of the $X^T$ energy is plotted as black squares (left axis) and the alloy dependence of the $X^0$-$X^T$ binding energy is plotted as red circles (right axis). Data was taken at 5 K with 1.96 eV excitation. Excitation and collection were done with right circularly polarized light (σ+). The error bars are equal to one standard deviation.

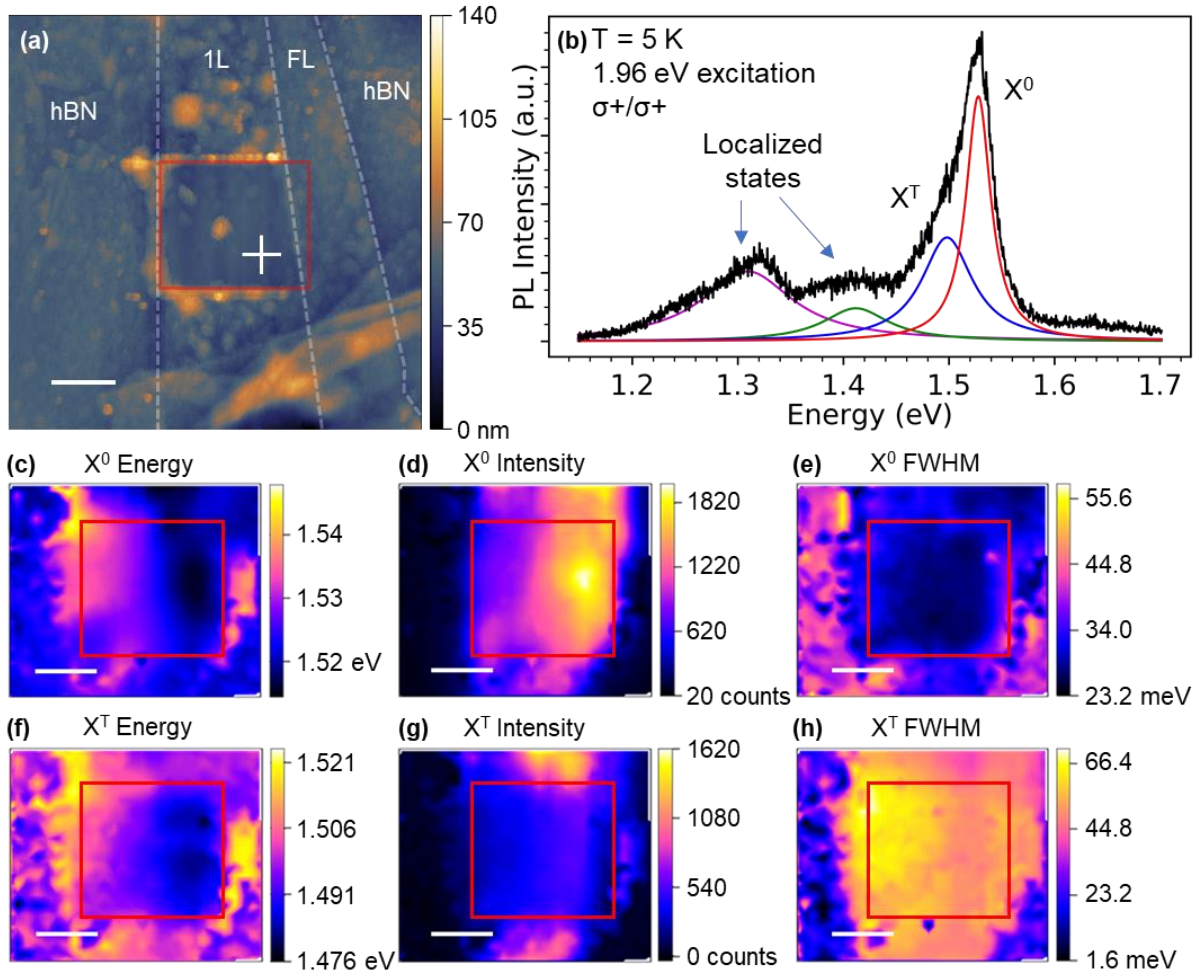

**Supplementary Figure 6 | Nano-squeegeeing of van der Waals heterostructures and PL spatial mapping. (a)** Atomic force microscope image of an hBN-encapsulated x = 0.33 sample that has been nano-squeegeed (region outlined with a red box) to clean the heterostructure's interfaces. Monolayer (1L) and few-layer (FL) regions have been labeled and outlined with white dashed lines to distinguish from areas with only top and bottom layers of hBN and no TMD in between. Residues removed from the heterostructure's interfaces are gathered around the edges of the nano-squeegeed region. **(b)** Low-temperature (5 K) PL spectrum taken with 1.96 eV excitation at the location marked by the white crosshairs in panel **(a)**. The sample is excited with right circularly polarized light σ+ and re-emitted σ+ light is collected. Lorentzian fits to the neutral exciton ($X^0$), trion ($X^T$), and localized states are shown. PL spatial mapping of the **(c)** $X^0$ energy, **(d)** intensity, and **(e)** full width at half maximum (FWHM), as well as the **(f)** $X^T$ energy, **(g)** intensity, and **(h)** FWHM. The nano-squeegeed region is outlined with a red box in panels **(c)** - **(h)** as well. All scale bars are 3 μm.

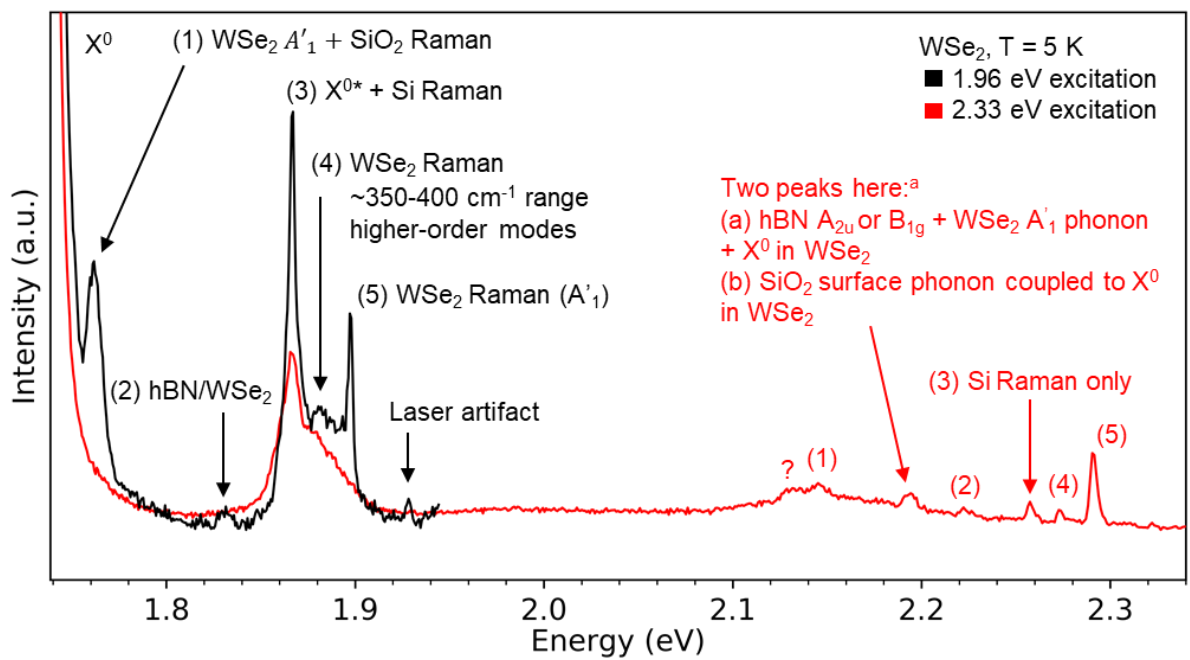

**Supplementary Figure 7 | Zoom-in of exciton and exciton-phonon complexes in WSe$_2$ at energies *above* X$^0$ that occur at 5 K.** Here, measurements are done with 1.96 eV excitation (black curve) and 2.33 nm excitation (red curve) to determine whether the features are related to optical transitions or Raman scattering. Peaks common to both measurements are labeled with numbers (1) - (5). The peak (3) X$^{0*}$ feature originating from the *2s* state of the A exciton does not shift with excitation energy, although it appears to decrease in intensity with 2.33 eV excitation since it no longer overlaps with the Si Raman peak as it does under 1.96 eV excitation. The peaks (1) - (5) that are related to Raman scattering shift in energy by the same amount as the change in excitation energy. The black curve ends just above 1.92 eV due to the collection cutoff filter. The superscript a refers to assignments made in Ref. 5.

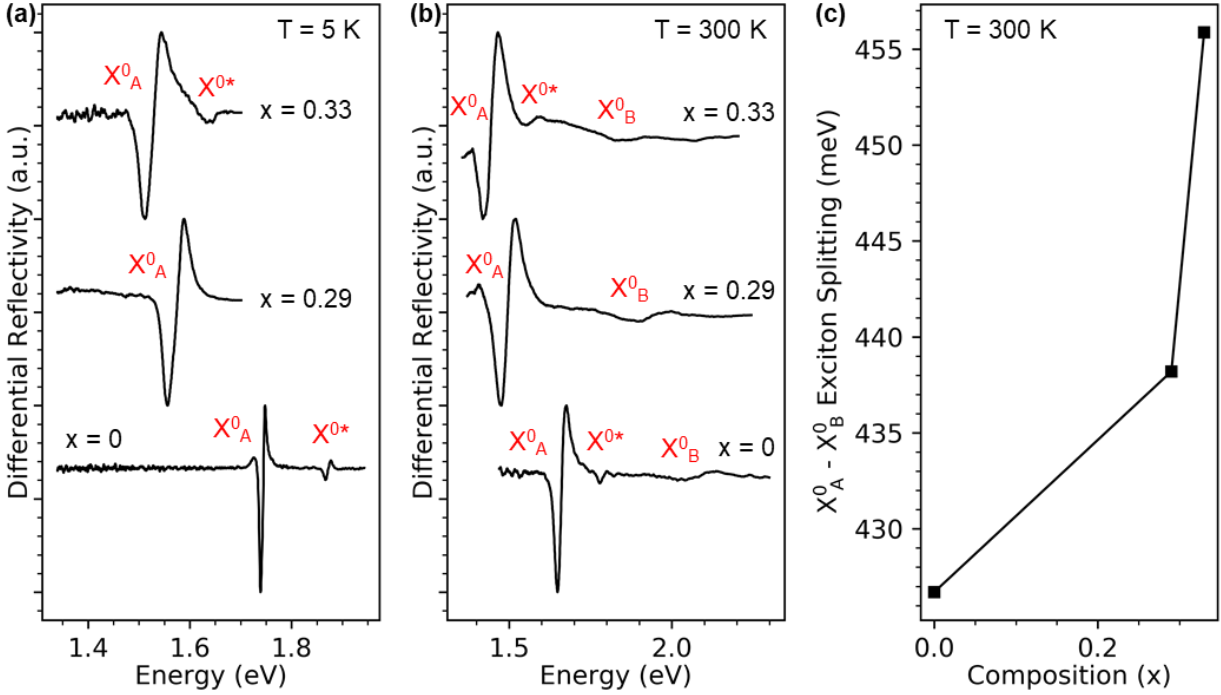

**Supplementary Figure 8 | Differential reflectivity measurements of select WSe$_{2(1-x)}$Te$_{2x}$ alloys.** Measurements are taken at **(a)** 5 K and **(b)** 300 K. Prior to differentiation, reflectivity measurements were processed using the equation $\frac{R_{TMD}-R_{Substrate}}{R_{TMD}-Dark}$, where $R_{TMD}$ and $R_{Substrate}$ are reflectance measurements taken on the encapsulated TMD alloy and the top and bottom hBN sandwich atop the SiO$_2$/Si substrate without the TMD alloy in between, respectively. $Dark$ refers to a dark scan to remove any background effects. Spectra are cleaned using a Savitzky-Golay filter. **(c)** Energy splitting at 300 K of the A and B states of X$^0$ plotted against alloy composition x.

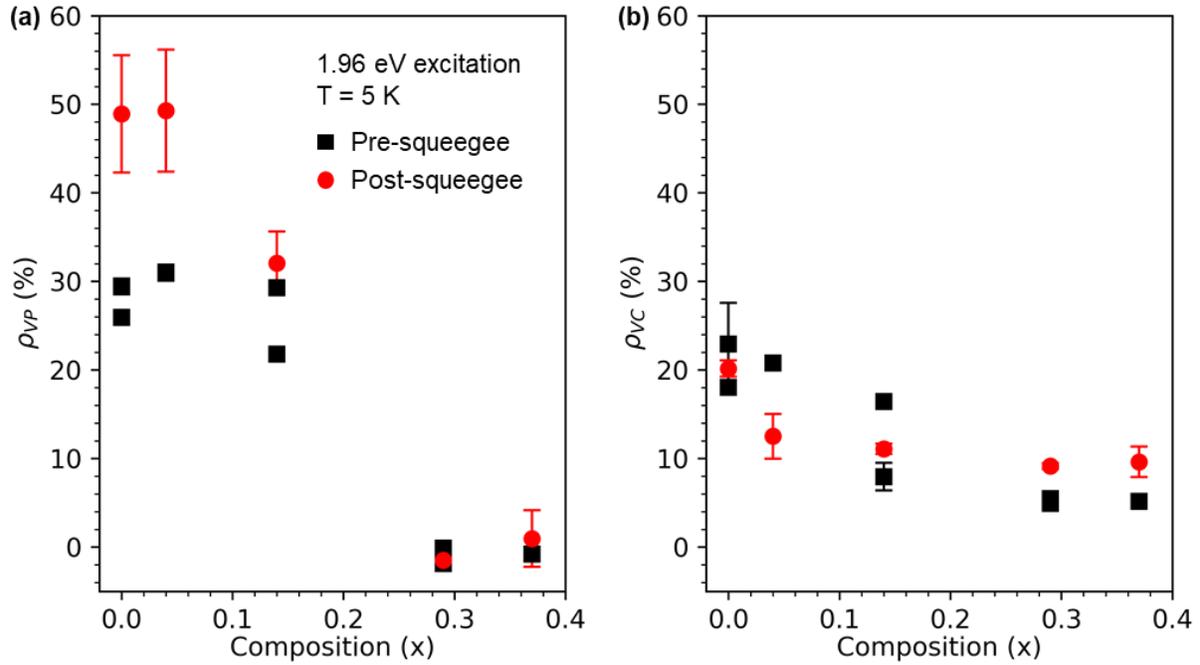

**Supplementary Figure 9 | Valley phenomena in 1H-WSe$_{2(1-x)}$Te$_{2x}$ before and after nano-squeegeeing.** Comparison of **(a)** $\rho_{VP}$ and **(b)** $\rho_{VC}$ for X$^0$ in samples pre- and post-squeegeeing (black squares and red circles, respectively). The nano-squeegeeing process improved $\rho_{VP}$ while having very little effect on $\rho_{VC}$. The error bars in both panels are equal to one standard deviation.

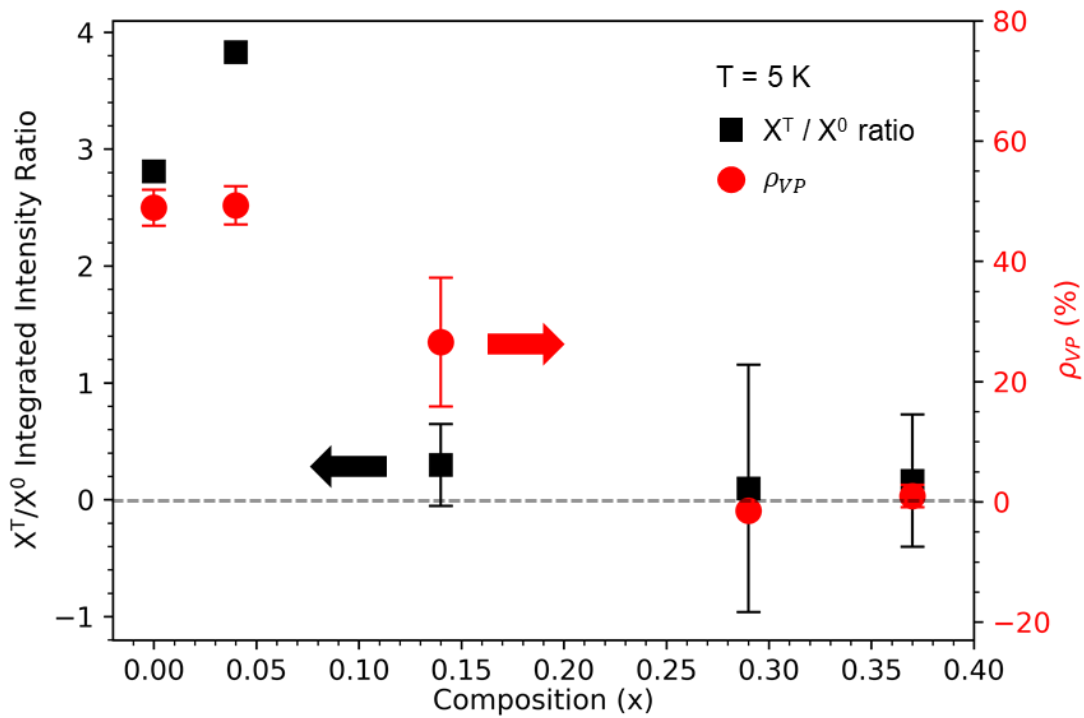

**Supplementary Figure 10 | Alloy-dependent ratio of the trion ($X^T$) integrated intensity to the neutral exciton ($X^0$) integrated intensity at 5 K compared to the alloy dependence of $\boldsymbol{\rho_{VP}}$.** The $X^T$ / $X^0$ ratio values are marked by black squares and correspond to the left axis. The $\rho_{VP}$ values are marked by red circles and correspond to the right axis. Measurements are done with 1.96 eV excitation at 5 K. The error bars are equal to one standard deviation.

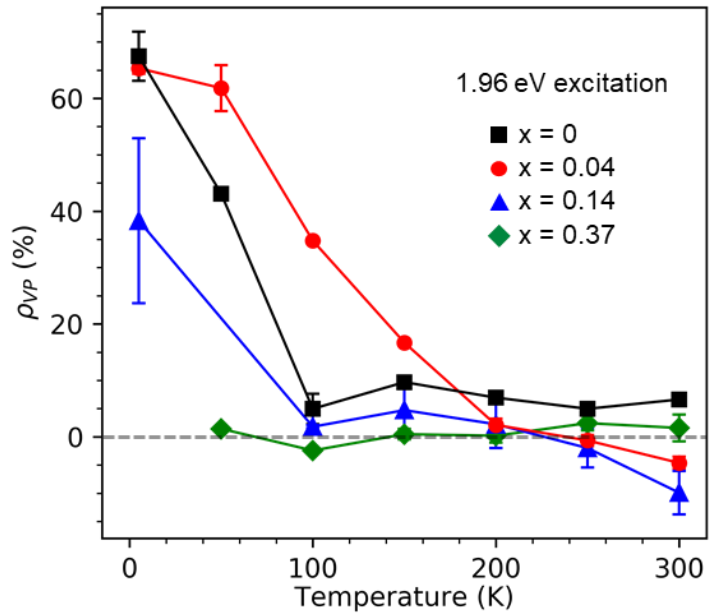

**Supplementary Figure 11 | Temperature dependence of $\rho_{VP}$ for $X^T$ in 1H-WSe$_{2(1-x)}$Te$_{2x}$ alloys.** $\rho_{VP}$ is found to be sustained at higher temperatures for the x = 0.04 than for pure WSe$_2$ (x = 0). Measurements are done with 1.96 eV excitation. The error bars are equal to one standard deviation.

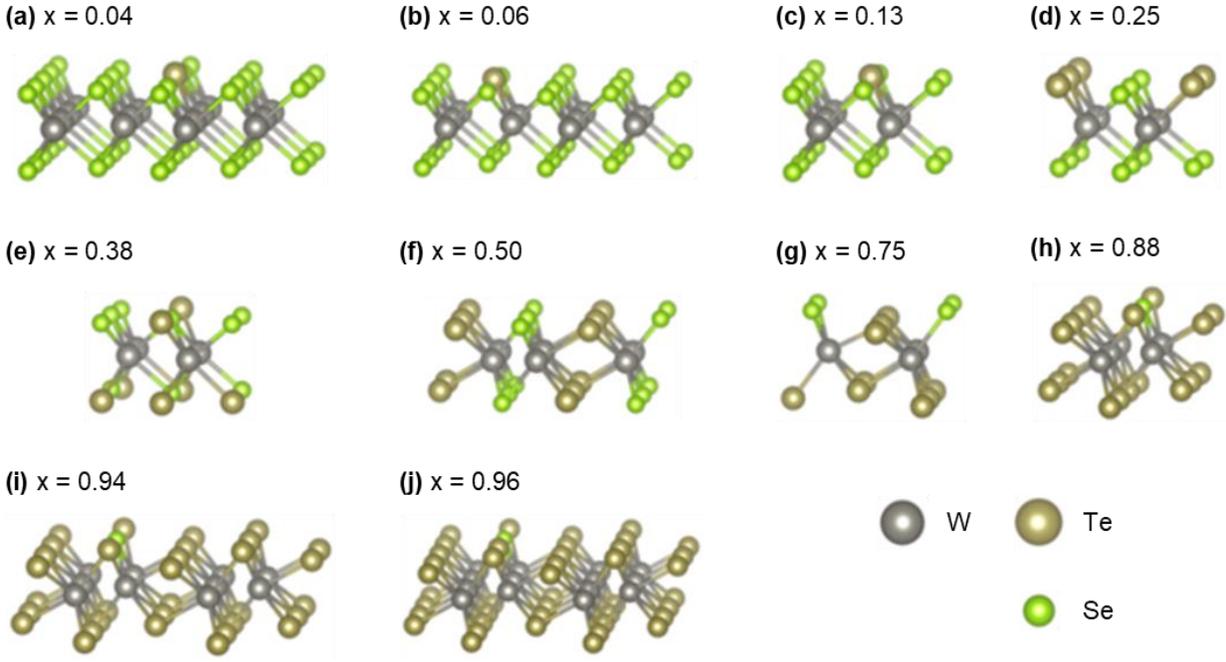

**Supplementary Figure 12 | Ground state structure of each WSe$_{2(1-x)}$Te$_{2x}$ alloy concentration investigated using density functional theory.**

**Supplementary Note 1**

We use PL spatial mapping of an x = 0.33 sample to explore spectral changes across the alloy (**Supplementary Fig. S6**). As with all of the other samples used in this study, the sample was mechanically exfoliated and encapsulated in an hBN heterostructure on a SiO$_2$/Si substrate using the dry-stamping method with a polydimethylsiloxane polymer.[3] An atomic force microscope (AFM) image of the sample can be seen in **Supplementary Fig. S6a**. We have used the nano-squeegee method[4] to ensure that residues from the encapsulation process have been removed from the areas of interest (red box in **Supplementary Fig. S6a**). Looking closely at the AFM image, it can been seen that not only have impurities been primarily pushed to the edge of the nano-squeegeed region, marked by orange colors indicating a higher height profile where residues have gathered, but some impurities faintly remain in the center and to the left half of cleaned region. A PL measurement taken at 5 K from the location of the white crosshair in **Supplementary Fig. S6a** is pictured in **Supplementary Fig. S6b**. This spectrum shows the typical features of the alloys, specifically the neutral exciton ($X^0$), the trion ($X^T$), and localized states, which in this case have been fit with Lorentzians. Spatial resolution of the energies, intensities, and full widths at half maxima (FWHM) of $X^0$ and $X^T$ at 5 K are shown in **Supplementary Figs. S6c-S6h**. We focus primarily on the nano-squeegeed region outlined with a red box. In this area, these measurements indicate that while the spatial variations of the FWHMs for $X^0$ and $X^T$ were miniscule, they were sharpest along the right side of the nano-squeegeed region. The areas of sharpest FWHMs hosted the greatest intensities of $X^0$ and $X^T$, as well as a uniform redshift of the peaks' energies.

**Supplementary Note 2**

We also find evidence for excitons and exciton-phonon complexes *above* $X^0$ that can be clearly seen in **Supplementary Fig. S7**. The lowest energy feature, labeled $WSe_2$ $A'_1$ + $SiO_2$ Raman, occurs at 1.76 eV with 1.96 eV excitation and shifts with laser energy for 2.33 eV excitation. The energy-dependent position of this feature suggests that this peak has a phonon origin, and was determined by a prior photoluminescence excitation study of different configurations of $hBN/WSe_2$ encapsulations on $SiO_2/Si$ and sapphire substrates to be a combination of the WSe2 $A'_1$ Raman mode and a $SiO_2$ surface phonon.[5] We measure another feature at 1.83 eV with 1.96 eV excitation, labeled $hBN/WSe_2$, that shifts with laser energy. As evidenced in Refs. 5–7, we attribute this feature to exciton-phonon coupling between hBN and $WSe_2$ that results in activation of the hBN IR-active $A_{2u}$ ZO mode or the hBN Raman- and IR-silent $B_{1g}$ ZO mode. On the other hand, the $X^{0*}$ peak measured at 1.87 eV does not shift with laser energy, although changing excitation energy results in a decrease in its intensity since this feature no longer rides the Si Raman peak background with which it overlaps when excited with 1.96 eV excitation. The presence of this feature in reflectance measurements at 5 K and 300 K indicates that $X^{0*}$ results from an optical transition (**Supplementary Fig. S8**). Assignment of this feature to the B exciton is ruled out since the $X^0$-$X^{0*}$ splitting at 5 K is ≈110 meV, which is much lower than the A-B exciton splitting of ≈400 meV,[8] and also both the $X^{0*}$ peak and the transition from the B exciton can be seen together in 300 K reflectance measurements (**Supplementary Fig. S8**). We suggest the $X^{0*}$ feature is the *2s* excited state of $X^0$,[9] in agreement with prior studies of hBN-encapsulated $WSe_2$ and $WS_{0.6}Se_{1.4}$.[10–13] We note that there is also a higher energy shoulder accompanying $X^{0*}$ that is independent of laser energy and is thought to related to the *3s* excited state of $X^0$.[13] The $X^{0*}$ feature is also observed in 300 K reflectance measurements of the x = 0.33 alloy (**Supplementary Fig.**

**S8**). In the alloy, $X^{0*}$ is ≈120 meV above $X^0$, suggesting only a small increase in the *1s - 2s* energy splitting with Te incorporation. Reflectance data illustrating the alloy dependence of the A-B exciton splitting at 300 K is also presented in **Supplementary Fig. S8**. This valence band spin-orbit splitting increases with Te composition as expected from prior DFT calculations.[14]

## Supplementary References

1.  Brothers, E. N., Izmaylov, A. F., Normand, J. O., Barone, V. & Scuseria, G. E. Accurate solid-state band gaps via screened hybrid electronic structure calculations. *J. Chem. Phys.* **129**, 011102 (2008).

2.  Ramasubramaniam, A. Large excitonic effects in monolayers of molybdenum and tungsten dichalcogenides. *Phys. Rev. B* **86**, 115409 (2012).

3.  Castellanos-Gomez, A. *et al.* Deterministic transfer of two-dimensional materials by all-dry viscoelastic stamping. *2D Mater.* **1**, 011002 (2014).

4.  Rosenberger, M. R. *et al.* Nano-"Squeegee" for the Creation of Clean 2D Material Interfaces. *ACS Appl. Mater. Interfaces* **10**, 10379–10387 (2018).

5.  Chow, C. M. *et al.* Unusual Exciton–Phonon Interactions at van der Waals Engineered Interfaces. *Nano Lett.* **17**, 1194–1199 (2017).

6.  Jin, C. *et al.* Interlayer electron–phonon coupling in $WSe_2$/hBN heterostructures. *Nat. Phys.* **13**, 127–131 (2017).

7.  Du, L. *et al.* Strongly enhanced exciton-phonon coupling in two-dimensional $WSe_2$. *Phys. Rev. B* **97**, 235145 (2018).

8.  Arora, A. *et al.* Excitonic resonances in thin films of $WSe_2$ : from monolayer to bulk material. *Nanoscale* **7**, 10421–10429 (2015).

9.  Wang, G. *et al.* Giant Enhancement of the Optical Second-Harmonic Emission of $WSe_2$ Monolayers by Laser Excitation at Exciton Resonances. *Phys. Rev. Lett.* **114**, 1–6 (2015).

10. Courtade, E. *et al.* Charged excitons in monolayer $WSe_2$: Experiment and theory. *Phys. Rev. B* **96**, 1–12 (2017).

11. Meng, Y. *et al.* Excitonic Complexes and Emerging Interlayer Electron–Phonon Coupling in BN Encapsulated Monolayer Semiconductor Alloy: $WS_{0.6}Se_{1.4}$. *Nano Lett.* **19**, 299–307 (2019).

12. Chen, S.-Y. *et al.* Superior Valley Polarization and Coherence of 2s Excitons in Monolayer $WSe_2$. *Phys. Rev. Lett.* **120**, 046402 (2018).

13. Manca, M. *et al.* Enabling valley selective exciton scattering in monolayer $WSe_2$ through upconversion. *Nat. Commun.* **8**, 14927 (2017).

14. Kang, J., Tongay, S., Zhou, J., Li, J. & Wu, J. Band offsets and heterostructures of two-dimensional semiconductors. *Appl. Phys. Lett.* **102**, 012111 (2013).